# Mikhail Vasil'evich Lomonosov

## *Discourse on Atmospheric Phenomena Originating from Electrical Force by Mikhail Lomonosov* [1]

*Oratio De Meteoris Vi Electrica Ortis, Autore Michaele Lomonosow Habita*

*Слово о явлениях воздушных, от электрической силы происходящих, предложенное от Михайла Ломоносова*

Among the ancient poets it was a custom, listeners, to start their poems with an invocation of the gods or with praise for heroes assembled among the gods so that their words would acquire more beauty and force; I considered it well to follow this example to begin this discourse. Coming to the formulation of the matter which is not only very difficult by itself and connected with countless obstacles but, more than that, may appear more formidable than before because of the sudden overthrow of the industrious accomplice of our zealousness[2], and coming to the clarification of this obscurity which, as I think, was introduced into your thoughts by this, confused lot, I must have greater fertility of ingenuity, a more delicate penetration into reasoning, and a more luxuriant richness of word than you can expect from me. Thus, so that my discourse would acquire importance and force and so that a lovely radiance should the subject of the suggested topic out of the darkness, I will employ the name of a hero the sole recollection of whom calls forth attention and reverence in all peoples and tongues. The works of Peter the Great are preached by the lips of all the mankind under the sun, and through the whole realm of the Russian autocracy the narration of these works arouses importance in the government councils, and sainthood in friendly conversations. For this reason, not only we recall here the majesty of such a great name with reverence, as my discourse needs the force and importance, but also the testimony of the grateful hearts of this whole assembly is in truth owing to its founder. For among the numerous great works of the great sovereign, this shrine of sciences in our motherland[3], founded by his incredible and almost divine great wisdom, was his main forethought. There are no any doubts about this for those whose impartial reasoning values the immense worth of wide extension of science for the enlightenment of the public, for those who saw themselves or were convinced by the greatness of fame and were astonished at the greater zealousness of the late sovereign to learn the teachings and personally propagate them in the motherland. For when the monarch, who was destined for great deeds, intended to set new troops against the enemy, to occupy a sea with a new fleet, to increase the sanctity of justice by a new majesty of laws to strengthen cities with new walls, to encourage corporations of merchants and diligence in the arts by new edicts and liberties, to correct the manners of all the subjects by a word, or, that the whole motherland be born again, then he clearly perceived that, without the use of mathematics, it was impossible to strengthen regiments and cities securely or to build, and launch ships safely; it was impossible to prepare weapons and fire-breathing machines and to prepare medicine for those wounded in battlefield combat without physics; that it was impossible to introduce laws, rules for the courts, or uprightness of manners without studying philosophy and rhetoric; and that, in a word, it was impossible to acquire an expedient defense for the state in time of war or ornamentation in time of peace without the help of science. For this reason, not only did him summon people known in all branches of science and arts from foreign shores to Russia with great rewards and with a tender and courteous reception; not only did he send selected youths like a multitude of hive-bees throughout all European cities and states renowned for their academies, gymnasia, military schools and artists' works, but he himself, the leader and public example for all burning with the search for knowledge, repeatedly absented himself from his motherland and travelled in Germany, France, England, and Holland, which was more than the customary procedure for other sovereigns. Was there one society of scientific people which he avoided or did not grace with his presence? By no means! He himself did not refuse to be entered in their number[4]. Was there somewhere a magnificent collection of splendid things, or a rich library, or works of respected arts which he would not see and about which he did not ask and did not observe with his honorable glance. Was there than a man known by the greatness of his teaching whom this great guest did not visit or whom he did not decorated with favor after a delighting conversation? What great outlays did he make to acquire valuable things produced by the manifold cleverness of nature and art which he thought would be suitable for the advancement of science in his motherland! And what rewards he promised if someone could tell him some great or new knowledge in the study of nature or art or promised to invent it! Although there are many eye witnesses of all of that present here, moreover, the many machines constructed by the untiring hand of the most august artist also bear witness; the majestic ships and the sturdy fortresses and wharves, the plans and construction of which were done rapidly and safely by his enterprise and under his command bear witness; the military and civilian schools established under his solicitude bear witness; this Academy of Sciences is a witness, furnished with so many thousands of books and with so many natural and artificial marvels and founded by the convocation together of men famous in all branches of learning; and finally, those very tools which are suitable for performing various mathematical computations and which followed him on all his trips also bear witness. For when he covered the waves of the Sea of Azov, of the-White Sea, the Baltic Sea, and the Caspian Sea with ships; when the conqueror and defender led his troops through Livonia, Finland, Poland, Pomerania, Prussia, Denmark, and Sweden; when he crossed the Steppes of the Danube and the burning Persian deserts, he had with him everywhere those scholarly people as his tools. From all this it is clear that he had to rely on all types of teachings for such great deeds; and they could not have been put to such a great advantage by anyone other than himself. Thus, when the use of science is very extensive, as per example of Peter the Great, not only in the good management of the state but in the renovation of it, for that reason we have to be assured by this true proof that those people who strive to discover secrets of nature via sacrificing labors or even by bold bravery should not be regarded as rash but rather wise and great hearted, and the studies of nature should not be abandoned even though some are deprived of life by unexpected fate. The fate of Pliny buried in the burning ashes of Mount Vesuvius[5] did neither frighten scholars nor even diverted them from Vesuvius's roaring ebullience caused by the internal fire. To our days curious eyes gaze into the deep and poison-belching abyss. Thus, I do not think that those minds inquired in the studying nature got frightened by the unexpected death of our Richmann and will stop studying the laws of electricity in the air; instead I hope that they bent all their efforts, with proper caution, so that some way could be discovered whereby the health of mankind could be shielded from those mortal blows.

---

[1] Translated to English from Russian version of Ref. [1]; all the footnotes added by translator
[2] Reference to death of Georg Wilhelm Richmann killed by a bolt of lightning on August 6 (o.s. July 26), 1753 while trying to read an electrical indicator on of his original design ("electric gnomon") [3, 4]. For many years Richmann (1711-1753) and Lomonosov (1711-1765) collaborated on the studies of atmospheric electricity [5, 6].
[3] St.Petersburg Academy of Sciences was stablished by the decree of Peter the Great on February 2 (o.s. January 22), 1724.

---

[4] Peter the Great was elected to Paris Academy of Sciences in June 19, 1717

Therefore, I, discussing upon the electrical phenomena in the air, and you, dear listeners, have much less to fear from that phenomenon, since so many calamitous experiments have been made already and paralipsis of them would be contrary to the general good of mankind. More than this, my considerations, apart from the topic assumed for the presentation, contain much about the atmospheric movements in general; there is nothing more useful to mankind than that knowledge. What more could be given by the highest divinity and permitted to mortals than that they should be able to predict changes in the weather which indeed seem to be the most difficult and scarcely comprehensible? But God recompenses us in full for our labors; it is possible to acquire everything from him through toil; we see a very clear example of this in the prediction of the movements of heavenly bodies, which was a mystery for so many centuries.

For this reason, often when gazing at the sky in my free hours, not without regret does the thought come to mind that many chapters of natural science have been explained even in the smallest details, but that knowledge about the circulation of air is still enshrouded in great darkness which, if it had been raised to the same level of perfection, we would now view on the same level as the other factors; everyone can easily judge what a great acquisition this would have been then for the whole human society. Indeed many and almost countless observations of the movements and phenomena occurring in the air were made by examiners of nature not only throughout all Europe but also in other parts of the world and were reported to the scientific community in print, so that a real certainty in the prediction of weather could be hoped for, - if not the imperfection of the instruments invented for this work, the differences in circumstances, the unequal care of the observers, the great and irregular multitude of the observations, the power of various considerations, cares, various propositions of sharp minds would not lead to confusion, would not overwhelm and would not oppress. Thus, as the complete perfection of instruments, an accurate knowledge of the circumstances, due care of the observers, and a detailed disposition of the observations not only were lacking to all but even were desperately needed by many, for these reasons the atmospheric movements seemed to be observed by the physicists not so much for the purpose of explaining them but rather as fulfillment of a duty. In such situation, the best part of natural science was worn out and almost put to death. But the happiness of our era finally encouraged everyone to get a good faith in it and apply all possible effort and it was as if a certain standard had been raised and they had good faith in it and applied all possible care. The works of those people studying nature got inspiration from the heavens when it was unexpectedly clearly

---

[5] Pliny the Elder, Gaius Plinius Secundus (AD 23-79), Roman nature-philosopher, author of multi-volume *Natural History*, died in 79 AC during eruption of Vesuvius

revealed that that terrible deadly fire originating in the rumbling clouds was similar to electric sparks, which their vigilance had, in our days, taught them to derive from bodies. Hence the researchers of the natural secrets turned their thoughts and hearts to the consideration of atmospheric phenomena, and especially of electric phenomena. And I, following this from afar more by reasoning than by experiments will briefly set forth what successes I made, as far as the circumstance of time and your patience will allow.

As physicists know, electricity is excited in bodies by two means: friction and heat. The phenomena and the laws that are generated by the electrical force originating in the Nature's bosom, correspond completely with those shown in skillfully set experiments. But as the Nature in the conduct of its various affairs is liberal and lavish, while it is miserly and sparing in giving reasons for these affairs, and moreover, since the same effects must be ascribed to the same causes, for this reason there is no doubt that the same causes, i.e., friction and heat taken separately or together, are the causes of natural atmospheric electricity. But who doubts that the vapors which move about the air can be warmed by the sun and can rub against each other due to the flow of air? Perhaps only someone who is not sure about the solar rays and about the fluid nature of air could doubt that. Thus, it is very probable that atmospheric electricity can be caused by the heat and friction of vapors; the question must be investigated for this reason, to discover whether atmospheric electricity indeed comes about in this manner; in the first place by the heating of solar rays. We cannot talk so assuredly about the upper vapors as about the materials close to the earth surface; besides referring to Boyle's observations[6], it is possible to draw conclusions about the sun's rays from the properties of certain grasses, properties which they always have. I should avoid pointing to the sunflowers, which are better known by the verses of the ancient bards than by the assuredness of the writers of natural history for the fact that they follow the movement of the sun, which property is not always observed in them, however the marvelous conformity of other plants to the movement of the sun increases the semblance of truth in this. It is confirmed by the proof of daily experience that many grasses which have their leaves open all day long close them at sunset and open them again at sunrise. Thus, there is a basis here for the assumption that same happens to thin threads attached to an electric machine which, when excited by electricity, separate one from another and assume a conical form; moreover, they hang next to one another straight towards the ground. Furthermore, the probability increases with an examination of that welcome and marvelous effect of nature which astonishes us in the new American shrub called the sensitive mimosa[7]. For besides exhibiting similar changes during sunrise and sunset, the fact that the mimosa lets its leaves down and drawing together even from a touch of the hand as if, it seems, hinting that the application of a finger diminishes its electrical force, while disconnection returns it again and that the leaves rise and open up little by little. It is true that many raised to refute my conjecture; however reasons will be found with which the correctness will be permitted to dismiss the doubts. It seems to be not in conformity with the laws of electricity if we assume that electrical force is generated during the day in the shrubs mentioned above without the required electrical supports, i.e., without putting beneath resin, glass, or silk; also that the electric indicator[8] does not always show the electrical force when the sky is clear, the sun is burning, and the sensitive mimosa has its leaves open. The first objection can be answered by the fact that the nodes of the grasses sensing the presence of the sun are fatten with resinous material and act in place of the support; the second by the fact that electricity which is produced by natural heat is weaker than that generated artificially, and for this reason can be sensed only in the delicate constitution of certain grasses. Besides, experience confirms this opinion of mine by seemingly no so weak proof. Last August 3rd, at sunset, having placed the sensitive American mimosa grass on the table, l hooked it up with an electric machine. The leaves were already compressed and were drooping from frequent touches of the hands so that not a single sign of sensitivity was seen after repeated touching by the finger. But when the machine was started up and the electricity began to act in the sensitive mimosa, giving off sparks to the finger, then the leaves, although they were not open, however drooped after a touch much below the hand. As many repetitions of this experiment proved, not without pleasant surprise, that the sensitive mimosa is greatly activated by the excitation of electrical force and that its sensitivity has something to do with electricity.

---

[6] Reference to Robert Boyle, *Exercitationes de Atmosphaeris Corporum Consistentium…* (London, 1673) [7]
[7] Presumed to be Brazilian "shy mimosa", *mimosa pudica*

Many different experiments of this type for a better understanding of this verity could be made on grasses that are sensitive to sunrise and sunset; but the shortness of time available for presenting the other material of this discourse keeps me out from this.

There is no doubt whatsoever that the friction of vapors against air occurs and can generate electric power. Now we must consider whether this really takes place and how it comes about? In pondering this question the thought comes to mind that the friction of vapors must occur through direct collision of these vapors; and this head-on collision can result from nowhere but from oppositely-directed movements of air which contains these vapors. Air movements in the atmosphere are very frequent, and some regularly occur parallel to the earth's surface in different directions, driven by different breaths of the wind. However, there is no way of substantiating the fact that winds produce atmospheric electricity. For if a phenomenon does occur in the in the absence of a factor, and, on the contrary, in the presence or approach of it the phenomenon does not occur, so that factor can't be either the cause or effect of the phenomenon. Because of the lack of interconnection, the winds and the electric power are usually time separated by to a great extent. Clear and calm weather almost always precedes occurrence of clouds laden with lightning. Whirlwinds and sudden wind storms with thunder and lightning are undoubtedly generated by these clouds. On the other hand, when violent air currents blow over vast territories and often blow over one place in different directions, which is indicated by cloud movements, strong collisions and friction should occur between these wind currents, and consequently, during cloudy and windy weather, lightning should flash, thunder should roar, or at least there should be some sign on the electric indicator if these atmospheric motions are the source of electricity generated in the atmosphere. But this hardly ever occurs, and thus by indubitable proof we are convinced that all the motions of air parallel to the horizon, i.e., the winds no matter from which direction they come, are not the reason and basis of thunder and lightning. But, one says, the motions of air are necessary for the collision and for-the electric friction of vapors; and except for the winds there is nothing that reaches our senses. That is the very truth. However, the action of electric sparks and its similarity with lightning was not investigated for so many centuries. "Nature does not commit all its divine acts together - reflects Seneca[9] - We expect ourselves to be consecrated already when we are still only in the entrance. These sacraments are open to all not without examination, but they are removed and shut up in the inner sanctuary. Much is left to future centuries when our memory will have disappeared; of which some will be discovered by the present time and some by a future time after us; great works are being born over a long time, and especially if they were ceased". We rejoice about this prediction of the noble philosopher which came about in our time, and apart from the other great discoveries we marvel at electric power which, when it was discovered to be similar to lightning, exceeded the astonishment of all.

---

[8] Electric indicator or "electrical gnomon" – electrometer built by Georg W. Richman early in 1745 [8]. It consisted of vertical metal rod and silk thread connected to its top. Electric charges on the rod resulted in deflection of the thread and angular deviation measured over a scale of an arc gave a measure of the electricity.

Those who discovered so many secrets concealed in nature by effort, or even sometimes unintentionally, attained truly great and just fame, and following in their footsteps should be considered not as a least praise. For this reason, I hope to merit certain gratitude when I find that there are air motions about which there is no clear and detailed knowledge, as far as I know, or at least that there is not as much detailed explanation of them as they are entitled to, and when I conclude that on a clear noon there are air motions perpendicular to the horizon which are the source and beginning not only of rumbling atmospheric electricity but also of many other phenomena within and outside the atmosphere. In order to present this in a tolerable fashion, I will follow the path which my reasoning has held to in the experimentation and discovery of those motions and phenomena.

I was often amazed when I noticed that during the winter throughout a dissolution of air in which snow melts, suddenly severe frosts come which, in several hours, lower the mercury in the thermometer from three or five degrees above freezing to thirty degrees below freezing and in the same time cover an area of more than one hundred miles around, about which it is rather possible to be convinced by rumors. Then, in comparing this with the winters of 1709 and 1740 which raged almost throughout all Europe, I was even more amazed and became even more desirous of finding the reason for such a sharp change. The fact that thaws almost always occur during overcast weather with a blowing and rapid flow wind turns out to be the most wondrous thing of all; frost, on the contrary, starts to show its severity after the winds have abated and the sky is clear. The fact that thaws are caused from the origination and nature of winds which blow gently is rather clear. For it is known from everyday observations that the severity of frost is mollified by storms blowing in from the depths of the sea. Thus, the winds blowing in St. Petersburg from the equatorial west, around the city of Arkhangelsk from the north and the south west, in Okhotsk on the shores of the Penzhinskaya Bay from the equatorial and north east tame the ferocity of the winter's cold, bringing in rainy weather. For this reason, Great Britain, across which no other winds except sea winds call blow, experiences a milder (shorter) winter than do other European territories at the same latitudes. In a similar manner, severe frosts rarely occur in Kamchatka which is subjected to sea winds from the south, east, and west, and enclosed by high mountains to the north; meanwhile the lands lying in Siberia at the sane latitude as Kamchatka suffer penetrating frost throughout the whole winter and rarely have thaws. For the enormously great distance of the open seas rising to the European and Asiatic shores, the Arctic ocean covered with permanent ice, the majestic mountains white with snow to the south which separate Siberia from India, in winter cut off the warm breeze from all sides. One does not have to marvel at the fact that the wind from the open sea during the winter bring a thaw with them over the land; because it has been investigated by experiments, that even under ice the sea water does not cool below the level of freezing to which its liquidity attests; and that when sea water is set out in a vessel during a frost, which forces the mercury to fall more than three degrees below the freezing level, then the water turns into ice. In agreement with judicious reasoning is the fact that the liquidity of sea water and the temperature are maintained above or around the freezing level for the great expanse of sea and for the underground warmth which breathes through the sea depth. Therefore, in the winter the open seas and those free from ice transport more heat to the air lying over them than does the land mass blocked under the frozen earth's crust and buried under deep snows through which the path is closed for the breathing of the underground heat.

---

[9] Here – citation from Lucius Annaeus Seneca: *Naturalium Quaestionum*, Liber VII, cap. 30-31

Thus, from observations and from the properties of the thing itself, it is clear what happens when there is a wind blowing from the sea to the dry land during the winter; the question remained to be considered is what occurs when the sea winds stopped blowing? Considering that, I attract attention to the difference in the heat and density between the lower air and that circulating above. There is more heat near the Earth's surface than aloft, or, according to the general concept, it can be said that in winter the cold is more intense above the clouds than below them at the earth's surface, and this is a true statement that has been shown by logic, studied by experimentation, and which agrees with atmospheric phenomena. In the first place, dense matter accepts more heat than more disperse matter of the same type. And this is a strong proof that the upper part of the atmosphere is heated much less by the sun than the lower atmosphere, while the middle part is affected depending on the height and other circumstances. Furthermore, the earth's surface heated by the sun and the rays reflected from the surface have a greater effect in the lower atmosphere than in the middle and upper atmosphere. The correctness of this reasoning is confirmed by several observations. Summer hail and the peaks of high snow-capped mountains right before the eyes show and convince us that not far above our heads, even in the midst of summer, the great severity of winter always prevails. Here I remember with delight the trials of famous men who, in order to investigate the nature of vast space, having sailed across the sea and overcome wide deserts, arrived in beautiful Peruvian localities[10]. Not being detained there by the pleasures of the meadows or gardens, nor reveling long in the gentleness of the heavens, but mounting the rocky peaks of the high mountains to measure the globe, they endured great cold and shed much sweat. By their lengthy and serious experiments and by their accurate calculations it was shown that at a certain determined height severe and continuous cold reigns throughout the atmosphere and covers high mountain peaks with eternal snow. The, degree of cold which extends from sea level to the snow line of the atmosphere decreases with distance from the equator, and finally disappears at the poles since the snow line merges with the ocean surface. It is clear from the following just how great the power of the cold is in that part of the atmosphere. In the first place, the famous globe measurers

---

[10] Reference to the French Academy's expedition to Peru to measure the length of a meridian degree under command of C.M. De LaCondamine, L.Godin and P.Bouguer (1735-1745) [9]

withstood, at a point in the middle atmosphere above the snow line, a severe cold of such a degree that hardly occurs in our lands even in the dead of winter. If such a cold prevails above the equator itself, it is easy to imagine how severe is the cold in our latitudes at around the same height. This reasoning is confirmed by a diligent examination of hail. Undoubtedly, the snow nucleus, which each hailstone contains under ice shell, is produced in the cold snowy portion of the atmosphere; the ice shells grow during the fall of hailstones through various layers of rain clouds, freezing because of the intense cold within the snow nuclei. It may seem impossible to the one considering it that falling hail could grow to a size the diameter of one's finger by new freezing of the water vapor in the very short fall time and at the rate of friction with the air, but this really happens and gives evidence of the terrible cold which is generated aloft in the snow nucleus of the descending hail. And if this takes place in summer, what does happen in winter? That is evidenced by places in Siberia in our same latitude but higher above sea level. The city Yeniseisk which lies more than 1500 versts[11] from the mouth of the river from which it got its name[12] is about 100 sagenes[13] above sea level, if in general it is assumed that the ratio of the fall to the length of the course is 1 to 7000, that is half a foot for each verst. Often the cold is so severe in Yeniseisk that the mercury in the thermometer drops to -131° below the freezing level[14]. Thus, there is no doubt that during the winter an equally severe cold exists above us at a corresponding elevation or, let it be at a higher elevation. In this situation, let us assume that after a sea wind the lower air has a temperature of four degrees above the freezing, and at a height of one verst, the temperature is the same as in Yeniseisk, so that there is a difference of 135 degrees. From my many experiments and from computation, it follows that the upper air in this case should be denser by one fourth than the lower air. Actually the density of the lower air increases due to the pressure of the upper air; for this reason, the density of the upper air at a height of 100 sagenes does not decrease by more than 1/48; and at 200 sagenes by 1/24, reckoning 15 sagenes to one barometric line. Hence it is clear that often the lower atmosphere is less dense and proportionally lighter than the upper atmosphere. The result of such condition of the atmosphere follows from aerometric rules and has been confirmed by examples. Earlier I explained the movement of air in mine shafts resulting from different densities[15], where for similar reasons at 50 sagenes or less there is often air current of such kind. Moreover, during the winter time in homes the warm air rises around the stoves and cold air settles around the windows, which can easily be discerned from the movement of smoke. So, can air at the height of 100 or 200 sagenes, exceeding by a good deal the weight of the lower air, be contained against natural laws? It descends and little by little mixes with the lower air, spreading a cruel cold over us. The denser air settles without detectable breathing, as that in one second it moves but a few inches or in two hours it descends 100 or 200 sagenes, contending with the air rising upwards. The indication, or better, the action of these movements in the air is very clearly demonstrated by the mixing of smoke rising from a stove pipe, because the smoky air which rises from a fire is always warmer and thinner than the surrounding air; thus, in summer time it rises to a considerable height and, having reached same degree of temperature as the surrounding, it ceases to rise. Therefore, in winter smoke should rise more rapidly and higher than in summer; however, something quite opposite to this repeatedly happens, and the smoke coming from a chimney goes downward rather than upward, so that the smoke cloud breaks up at the very chimney outlet and stretches from the housetops to the ground. That this is not due to cold and extraordinary density of air is clear from the fact that when the cold continues for several days, the smoke not only does not produce a cloud that reaches the ground but, ascending higher than is usual, forms columns of smoke like high trees in the still air. The second action of these motions is the clearness of the sky, for although much can be ascribed here to air density, the clouds separate, disperse, and disappear due to the ascending and descending motions of the air. Therefore, sudden frosts in winter are caused by the subsidence of the middle atmosphere towards us, and such advents of cold that begin without any wind action now should not appear as such a incomprehensible action.

---

[11] 1590 km

[12] Yenisei river

[13] 213 m

[14] About -87.3°C; here the temperature is in Lomonosov's thermometer scale in which the water freezing temperature is set 0° and the temperature of boiling water of 150°

[15] Reference to Lomonosov's dissertation *On Free Movement of Air, Observed in Mines* (1744, *Novi Commentarii*, V.I, St.Petersburg Academy of Sciences, 1750, in Latin)

Similar submersions of the middle atmosphere into the lower atmosphere must occur in summer as well, to which the disposed arrangement of air towards the lower rather well attests. Let us assume that the air which can produce hail in summer is at a height of 300 sagenes and is $50^o$ below the freezing point, which can be safely presumed from all considerations; at the same time in the lower atmosphere, near the ground, the air has been heated to $40^o$ or $50^o$ above freezing; then, according to my experiments and calculations the density of the upper air and the lower air will be in the ratio 6:5, while the lower layer will be compressed by the pressure of the upper atmosphere and become denser than the upper by about 1/10. Under these conditions, according to the inviolable laws of nature, the upper part of the atmosphere should descend into the lower part to a depth where, mixing with the warm air, equilibrium is reached. This flow of ascending and descending air should occur as often as the weight of the upper air exceeds the weight of the lower; moreover, the lower air should encounter the upper air and collide with it at various heights and in various directions, depending on height and differences in heat and density; finally, this should occur the better when the earth's surface having been heated by an intense summer hot spell, heats and expands the air lying above its surface, while the intense cold above the clouds compresses the middle portion of the atmosphere.

It is already sufficiently clear what kind of air motions, besides the blowing of the wind, can create electric friction; thus, it remains to be discovered whether there are materials in the air and how they act so that an electric force can be excited by the collisions of these materials. Two types of materials are required for this: the first, materials in which electricity is generated, and the second, materials which accept the generated electricity. Water, more than any other material, absorbs electricity and there is plenty of water circulating in the air, as indicated by abundant rains, which occur at that very time when the air shows that it contains electric force. Among the substances (bodies) in which electricity can be excited by friction, organic materials[16], which can be burn with a flame, play major role. Such particles are abundant in the atmosphere as can be easily proved. First, there is the imperceptible emission of vapors from substances, oxidation, and the rotting of plants and animals throughout the world; there is burning of material to protect our bodies from the cold, to prepare food and to produce various things artificially which are necessary to sustain life; moreover, there are fires which burn down houses, villages, cities and great forests; finally, there is the constant emission of smoke from volcanoes and frequent eruption of a bright flame which can conveniently express the absolutely enormous amount of organic combustible material that there is in the atmosphere. Secondly, the abundant growth of succulent grasses, which can set their roots in infertile sand, clearly shows that the leaves of these plants can take nourishment from the air, because the plant could not get that much resiniferous material from the dry sand. Thus, we have material of both kinds in the atmosphere which are capable of producing electric friction; now it remains to investigate the manner in which these materials meet, collide, and rub against each other.

---

[16] also can be translated as "oily/combustible materials/particles"

It is clear from accurate chemical experiments that volatile substances by the difference in their nature differ from one another in weight and rate of ascent, so that pure combustible vapors, rise higher than water vapors. They can always be distinguished by the differences in rates of ascent even in a small height such as chemical vessels have and, thus, there is no doubt that updrafts of combustible vapors will rise much higher in the free atmosphere and, separating themselves from the water vapors, will collect above the water vapors. There are two known types of fine combustible vapors: one combines freely with water and is called simple double-distilled alcohol; the other does not combine with water and has been called volatile oil by the chemists. The first, when it rises, combines with water particles in the clouds and hardly rises any higher than them; the second type avoids water vapors and rises above them, all of which agrees with the laws of nature.

Moreover, this agrees with daily observations, since we often see two or three rows of clouds at different heights, which have risen to these heights due to the differences in their lightness. It often happens that other vapors of an organic substance lie above several layers of clouds consisting of water vapors and these vapors will remain there as long as the atmospheric density remain in equilibrium. However, as soon as the lower air expands and becomes thinner due to heating, the cold and dense part of the atmosphere will be forced to descend, and the lower part will rise to take its place. I shall attempt to explain: as briefly as possible the phenomena of these changes to your mental eyes, in so far as you can understand from my discussion and, as you yourselves have seen, can remember.

When the upper atmosphere descends due to its greater weight, as it extends it does not settle evenly horizontally everywhere, but will create various air densities, depending on the position of the clouds and the irregularities of the earth's surface, and from the variations of the sun's rays. Thus, air will descend in places where it is denser and heavier - in the shadows of a mountain or of a high building or a cloud the air is denser; air will rise from places where the mountain slope faces the sun, or through cloud openings heated by incident rays. For this reason, when thunderstorm clouds rise before a rain, then the lower clouds for the most part move upwards or downwards like mounds, and shaggy-looking vapors extend toward the earth and curling eddies develop, dark openings form, and more than this, above these phenomena, the clear sky becomes covered with a dull, dark blue color. All these circumstances show that, upon descending, part of the middle atmosphere is filled with combustible vapors, and, for this reason, the clear sky is covered with this blue darkness which penetrates the lower clouds with its uneven submersion, and, passing through them, contends with the air it encounters. Due to the submerging of the upper vapors and the air rising from below, the clouds bend upward, which causes twisted and straight vapor braids, especially if a water cloud has been penetrated by combustible vapor.

At that time, the organic globules of combustible vapors which cannot combine with water vapor because of their different nature and which approach the properties of a solid due to their immeasurably small size, collide with a rapid head-on motion, rub against each other, and create electricity which expands itself throughout the cloud until it occupies the whole cloud. It may seem strange that such a frightful force is created by such tiny spheres, but your amazement will cease when you consider the countless number of these spheres and the infinite surface of water substance in the cloud, a surface produced by the separation of the substance into tiny particles. For it has been found by experiment that the greater the surface is of a body of derived electric force, the more electric force it will be able to take on. Iron, wrapped around with braid, repeatedly produced an intentional action from glass balls which are not very capable of producing electricity, an action which did not show itself acutely when touching those spheres. In a similar manner, large clouds, divided into tiny particles and in a confined space, take on an enormous electric force, exhibit terrible actions and disturb the mind with their inconceivable outcomes, the most important of which I intend to interpret here by the laws of electricity. However, before this I will attempt to explain the general phenomena of thunderstorm clouds on the basis of my theory in order to show its great possibilities.

First, it is generally known that clouds laden with thunder and lightning for the most part ascend after noon and assemble around three or four in the afternoon when the action of the sun in heating the air is the greatest[17]. These circumstances agree with my reasoning. For the more the lower atmosphere is heated up, the more capable the upper air will be to descend into it. That part of the atmosphere that feels less heat thins out less. This can be shown by the rise of mercury in a thermometer and by the fall of mercury in a barometer, when the two are observed simultaneously.

Furthermore, as everyone knows, hail often falls from thunderstorm clouds after a very hot spell. In fact, your tactile sense will tell you that the upper atmosphere is very cold during the approach of an electric cloud, and the effect of this or some part of it will even reach us.

When the sun's rays are blocked by clouds, the air in the shadow of those clouds should become cool and contract. Therefore, the air should move from the edges of the shadow toward the center. The same effect should follow the growth of falling raindrops, because the moist vapors, combining into water drops, absorb a great quantity of air. However, such movement of air into the middle of the shadow hardly ever occurs, but, as without doubt all of us have regularly observed, the opposite happens: when clouds laden with lightning approach, not only do they send out swift streams of air in front of them but, in passing by, they also send out strong lateral winds, and for the most part leave behind them a certain sereneness. Where does this great river of air

---

[17] Lomonosov's logbooks contain records of thunderstorms from 1744 to 1748, their dates and times, see *PSS* [8], vol.3, pp. 182-192

originates from? It comes from the fact that the lower atmosphere being squeezed by the pressure of the upper atmosphere, spreads out in all directions, and mostly rushes to the side where it finds the least resistance.

Furthermore, heavy rain showers occur during thunder and lightning, and, like an overflowing river, by the sudden fall of water, overturn large rocks, upset houses, and destroy fertile fields in the twinkling of an eye. What more than this transformation could prove the descent of the upper atmosphere into the lower? It descends, laden with vapors, combines with the clouds which are quite saturated with water and continues on down.

Finally, thunderstorms are more frequent and rage more dangerously in mountainous regions. Although generally known, it has been more strongly confirmed by the observation of Spanish natural scientists. In the Peruvian province of Quito which is surrounded on all sides by very high mountains which rise much higher than the snow line, terrible and dangerous thunderstorms shake not only the buildings but the mountains themselves and flood everything with torrential rains. These storms always occur in the afternoon, and are preceded by a clear and calm air in the morning, and such transformations happen over almost one quarter of the year. That agrees with my theory in such detail, as everyone can clearly see as soon as he considers that the air in mountainous regions is practically never in equilibrium. In places exposed to the sun the air rises, it sinks in the shadows and, thus, more easily draws down the colder and heavier part of the upper atmosphere, it accelerates its motion, excites a much more powerful electric force, and moves closer to the earth.

Judging by the agreement of so many events and phenomena, I hope that I can assume that my theory stands on firm ground. Therefore, putting aside further arguments which might be used to divert any doubts, I come to the atmospheric phenomena which occur in conjunction with thunder and which can be explained by the properties of electricity.

First, I intend to discuss the forms of lightning. Two common forms of lightning are observed. The first, with a red flash and crooked path, shoots out with thunder, storm, and rain; the other flashes near the horizon after sunset, it is pale and flashes with extensive radiance without thunder above the clouds, under calm and for the mostly clear air, behind the thin and sparse clouds. Three types of electric light are known. The first occurs in a spark with a crackling noise and often noticed with a crooked path and different colors depending on the material, especially when natural electricity was drawn from a cloud into a metallic rod. The second type is a hissing and cold flame which is encountered especially coming from sharpened metallic points when brought close to material, of the type I once saw in my room during great thunder and lightning, when it was 1 foot wide and three feet long, of commonly pale color and emitting a hissing sound but without the crackling noise. The third type is a pale and weak light which appears in very thin air or in places where there is no air at all, like over the mercury in a barometer, and it continues to flash uninterruptedly at regular time intervals even when the electric force disappears. Without a doubt, the electric sparks produced by experimentation which jump out with a crackling noise towards an approaching finger are of the same nature as thunder bolts. Evening lightnings, which are commonly called heat-lightning, apparently belong to the third group, since they occur in the thin air of the upper atmosphere and flash with a pale light after thunderstorm clouds, and, in addition, flash at equal intervals of time, which I repeatedly have noticed, reckoning some forty seconds between the flashes. The hissing light which comes from sharpened metals with that harmless flame must be considered separately; it sometimes shows itself above the heads of men, as in the Virgil songs about Lavinia[18], and also burns about the lances of Roman soldiers and around the iron maces of their heretogas. To the same group belong also he lights called Castor and Pollox[19] which appear with a hissing sound over sail yards after a storm, as witnessed by many.

As for the curves and the turns of lightning, I consider it very likely that it twists along a spiral line, therefore the curves, angles and rings appear depending on the different positions of the observers. This theory of the electric force occurring in the atmosphere and general experimentation are also in strong support of such notion. For when electricity is generated by the subsidence of the upper air, a cloud or air saturated with water particles bursts open and creates an action similar to the pouring of water into a drain; the organic vapors, descending through the water vapors, whirl in an eddy and direct the lightning towards a similar type of receptacle. Furthermore, a powerful electric force produced artificially gives off sparks which appear to be significantly bent. I have often seen sparks nearly an inch long jumping from iron filled with natural electricity to my finger and I can attest that they are part of a spiral line. The study of sparks was much more convenient when they occurred at the time of a strong thunderstorm cloud and continued almost incessantly, so that they crackled sharply towards the approaching finger in the manner of a source with such jolting that it was hardly tolerable for the whole hand. The first spark was always the strongest and struck with a more curved force.

The question of thunder bolts is still to be mentioned, which many have their doubts about; however, I do not dare to deny it altogether, since earthy material melted by a thunder bolt can produce it. These are my thoughts about the usual thunderstorm phenomena and circumstances. Those which occur less often and which produce more astonishment will follow.

It is well known, as not long ago it was discovered in Italy, that thunderclaps sometimes came out of cellars, and due to that, their cause was thought to be completely different from that of electric force. But this phenomenon relates to electric force in its entirety. For as soon as an electrified body approaches another which does not have that force within it, sparks leap out from both bodies to meet each other; however, the sparks are more powerful from the electrified body than from the one which has not yet received that force. In a similar way, an opposite spark, similar to lightning comes out of the cellars to meet the one coming from the cloud, as the cellars are made from solid and dump material suitable for the reception of derived electric force, and, moreover, dug deep into the ground and, for this reason, oppose an electric cloud with great force.

The narrations of ancient history and the reports of recent eye-witnesses tell us that fire falls from thunderstorm clouds to the earth. Because of its not so violent movements, such a flame should be considered as something separate and distinct from lightning, Thus, it is sufficiently clear here that the organic vapors are heaped while falling and light up, and then sink to the ground and this marvelous phenomenon corresponds to my reasoning.

---

[18] see verses 72-80 from the book VII of Virgil's *The Aeneid*
[19] so called "Saint Elmo's lights" appearing in the lower atmosphere, mostly over sharp objects

There is enough evidence both in the past and recently that thunder rumbled when the skies were clear, to the surprise of observers. Death of Professor Richmann occurred in similar circumstances. But that ceased to be surprising when we found out that, even during a clear sky, the air might have more of different types of vapors than it sometimes has even during an overcast sky.

The ancient writers left us notes of occurrence of rocky rains, and we read in more recent annalistic books about similar wonders, when during the ascent of storm clouds laden with thunder and lightning, rocks of frightening size were raised aloft, high trees were ripped up by the roots, and stone temples were overturned. This can be ascribed without difficulty to the attraction of electric force. For, having compared: the thunder claps and the great extent of electric force in the atmosphere with the electric sparks produced artificially and with the small range of action, it can be easily understood that such great bodies can be separated from the earth's surface and borne aloft into the air by this very strong and incomparably great force located in the vicinity.

Not only the earth but the seas feels the very great force of such attraction as well. "Typhon[20], a very great danger for seafarers, writes Pliny[21], descends something, having broken away from a cold cloud, twists and turns, increasing its fall and weight, and changes place by rapid rotation; breaking not only the sail-yards but also the vessels, turning them around. Then it reverberates with the striking force, bears the ravished bodies aloft and devours them on high. And when, heated up and breathing fire, it rages, being called Prester, and burns and dissipates everything it touches." Events similar to this experience are confirmed at the present time by sailors in a section of the ocean below a torrid zone who say that it is as if a column descends from a cloud to the surface of the sea which rises up to meet the column like a mound which boils up in the approach, and on the inside a lean cloud column swirls around in the manner of a screw. Finally, it disperses in a great torrent rain, and, with a frightful rumbling, as many carriages going suddenly over a street paved with stone, it pours out into the sea. All these phenomena and events which Pliny and others described can not only be explained by my proposed theory but in addition, firmly support it. The descent of the cloud column takes place due to the violent rush of the descending upper air, the screw-shaped cavity in the cloud column corresponds fully to the spiral path of lightning which was described earlier, the water mound which rises from the sea surface toward the cloud column, and also the fact that it (typhoon) tosses aloft broken up sail-yards and vessels,- all of these things are due to the attraction of a powerful electric force; the fire in the column is burning organic material. Then, after the cloud column comes in contact with the water mound and loses its electric force, having imparted it to the sea, then there is a great crackling sound and a drowning rain which digs up everything with its headlong motion. Here, I hope, they will ask how such an electric attraction occur without the usual thunder and lightning? My observations have answered this, through which I have found out that the air often contains a strong electric force without lightning and thunder. How this happens will be explained in the next part of this discussion, for now in the natural order of things it is necessary to explain the action of lightning, which is the most remarkable and miraculous of all.

---

[20] Typhon, also Typhoeus, Typhaon or Typhos – in antique mythology, a monstrous shaky giant symbolizing underworlds' volcano phenomena
[21] cited from Pliny the Elder's *Natural History*, vol. II, chapters 49-50.

Surprisingly, it turned out that objects around those struck by thunderbolts remain unharmed. But the astonishment ceased as soon as it was discovered that thunderbolts are subject to the rules of electricity and for this reason bodies of original electric force can be conveniently safe from the bolts of thunder. However, one question has been left up to the present time without explanation- that materials of the original force which are exposed to burning, such as silk, wax, and other materials similar to them, remain unharmed as distinct from the metals themselves which are melted by lightning. For although silk and wax are safe from thunder bolts, when the metal which is contained in them or is touching them is melted, then they should melt or burn before the metal becomes cool. The metal melted by a direct flame, and especially hard metal, has to take on such a degree of heat that it is incandescent for so long ant is so hot upon reverting to its solid state that it can destroy not only silk and wax, but also set fire to a tree and inflame a blaze. Well, what to do? Is it really possible to ascribe to lightning the very rapid force of being able to make a metal red hot and to cool it down within the same twinkling of an eye? But the basis of the contradiction. overcome by this, and the constant natural laws of the manufacturing and extinguishing of fire, refuted by this, thwart us. For' this reason, isn't it possible to assume then that metals deliquesce in cold without a real fire? By every reason! For as much fire as there is in lightning, not only can it not melt metal in the twinkling of an eye, but often it does not set fire to the driest tree by a powerful strike and only breaks and tears. The most powerful force of thunder manifests itself when the parts of a body that has been struck become divided from their mutual relationship by the terrible action. This happens with electric force produced artificially according to the measure of its smallness. For the fact that thread is repelled from a metal bar, that filings jump apart, that water flowing· from a narrow crack separates and breaks apart, and that rain shows itself of a conical form in its fall and by its tiny droplets clearly shows that electric force excited artificially drives even the smallest particles of a body from their mutual union and weakens the strength of their viscosity. It is clear from this that the attractiveness of the smallest particles gets weaker with the greater the electric force and the more capable the body is of taking on this electric force. Considering the immense natural force and the capability of metals by which they take on this force, it should not be very surprising that their particles are so repelled from one another that, in changing into a liquid state, the metal melts in that very twinkling of the eye in which the strike occurs, and after this acting cause is gone, the particles return to the compound of the former union in an unnoticeable time and all this sometimes takes place without the excitation of that kind of flame with which wax can liquefy. When explaining this surprising cold melting of the metals which have been struck by lightning in this manner, I saw it to be in agreement with nature, and I directed my thoughts to this and recalled my former works, then I saw, not without rejoicing, that my thoughts on the cause of heat[22] which had been communicated to the scientific community were in complete agreement with my theory. It is true that up to this time I still stand for the truth proven by many possible arguments that the cause of heat is in the internal movement of the very particles of the matter which compose the bodies, by which movement all the particles revolve around their centers. It follows from this that the extraneous matter which is contained in the minuscular gaps in between the particles of the bodies, can move without producing heat or fire. Electric matter which shows its rapid movement in cold bodies, even in ice, by violent sparks, confirms the truth of my thoughts, about which repeated experiments dispel all doubts. When the bodies are heated up, that is by making the rotation of the particles composing the bodies, then the force repelling from the center is strained, their attraction is weakened and solid bodies melt with the increase of fire. Thus, it is very probable that similar movements first stimulate the extraneous electric matter to produce other movements and other phenomena. For the heat and electric force are both produced by friction – the heat requires strong movements of larger particles, while electric force requires gentle friction to excitation of the finest particles, so that they will revolve around their centers. Thus, during the fast revolution of the particles of electric matter circulating in the minuscular chinks of metal, when it is excited by a thundering electric force and when the particles composing the metal stand quietly or move very little so that the heat of the metal increases very little or not at all, then a great force of the electric matter repelling from the center is produced in the chinks, it expands the particles, drives the particles from their bonds, and so weakens their viscosity so that the metal melts.

---

[22] Reference to Lomonosov's *Considerations on the Nature of Heat and Cold* written in 1744 and published in Latin in Vol. I *Novi Commentarii Academiae Scientarium Petropolitanae* (1750) – see also *PSS* [2], vol.II, pp.7-55.

Having explained these phenomena, I hope that I have satisfied your curiosity as much as possible by the thunderstorm theory, and for this reason I now turn to the part in which I will attempt to find convenient methods of avoiding death-dealing thunder bolts. I do not want by this undertaking, dear listeners, to arouse in you any indignation or any fear. For you know that God gave even the wild beasts the sense and strength to protect themselves, and to man he gave more than that – the penetrating reason to foresee and avert everything which might harm his life. Not only lightning bolts lash out at life from the bosom of abundant nature, but many other things as well: epidemics, floods, earthquakes, storms, which harm us no less and frighten us no less. And when we defend ourselves with medicines against plague, and with dikes against floods, and with strong foundations against earthquakes and against storms, and when we do not think that we have opposed the wrath of God by this very audacious reinforcement, for this reason what reason can see which would prohibit us from being delivered from thunder bolts? Do they consider as impudent and godless those, who for the sake of a contemptible profit cross the immeasurable seas raging with storm, knowing full well that the same thing might surely happen to them which formerly many of them, or at least their parents endured? By no means; but they are praised and even more than that they are commended by public praying to the protection of God. Should they consider them insolent and impious, those who for the common safety, for the glorification of the greatness and wisdom of God, pursue the greatness of his deeds in the nature of lightning and thunder? Not at all! It seems to me that they still use his special generosity, while obtaining the richest remuneration for their labors, that is, the discovery of such great natural wonders. We see his sanctuary laid open with the discovery of electrical phenomena in the atmosphere and we are summoned by a sign into the inner entrance - halls of nature! Shall we still stand at the threshold and be held back by the contradictions of groundless prejudice? By no means; but, on the contrary, however much has been given and permitted to us we shall not stop extending further, investigating everything which an intelligent eye can penetrate.

Thus, we shall examine, to the extent possible, the number, position, and acting force of clouds laden with thunderstorm electricity. At first the thought, comes to the mind of the one considering the question that sometimes there are many such clouds end sometimes only one. In the first case different changes occur according to the different cloud positions for either all the clouds receive electric force or only certain ones do, The first of these does not happen as often as the other as one can judge by the different heights of clouds, and when and if it happens, then there must be different degrees of electric force due to the difference in the heights of the clouds. Therefore, when there is electric force excited in a cloud which is located right near another one which has very little or no electric force, the electric force produces a crawling spark between the two clouds, that is, lightning and thunder. In a similar manner other clouds, communicating their force between one another, will flash and rumble among themselves for as long as the electric force in them holds out which can be exhausted in different ways. It very often happens that a sharp crackling of sparks from an iron arrow set up no higher than four sagenes[23] follows rapidly on the arrival of a thunderstorm cloud. It follows from this that the electric force in clouds extends to the surface of the earth and is taken on by objects of all types, and especially by those which have sharpened ends, through which the electric force is diminished and completely exhausted in the passage of time. This occurs especially when the extent of the electric action little by little becomes smaller and it gets weaker the further it spreads out from its cloud. On the other hand, when the limit of electric force directed towards the earth terminates sharply in the approach to the earth, so that the arrows that are set up do give no indication, then it happens that the cloud communicates its force to the ground by sparks and crackling, that is, by lightning and thunder, striking into those bodies which are either the closest of all or are of the greatest derived electric force. Hence, not without reasoning can we expect that those clouds are the most dangerous which do not exhibit the slightest electric indication between violent lightning and thunder, at the set-up arrow. It follows from this that, by comparing the declination of the thread from a metal bar[24] with the interval in time between the lightning flash and the thunder clap, it is impossible to determine reproach of lightning. In addition, it can then happen that the gap separating an electrified cloud from another nonelectrified cloud is directly above us and for this reason the spark which occurred between them communicates both the crackling of lightning and thunder almost simultaneously to our sight and hearing. Meanwhile, those who are located underneath the edges of both clouds on opposite sides away from the gap, hear the thunder later while having seen the lightning at the same time; and between themselves they can notice such a difference, that the current on the arrow under the edge of the electrified cloud, shows more force before the lightning than after the fire and, on the other hand, the person who stood under the weakly electrified or a non-electrified cloud after the bolt had struck sees an increase or the generation of the force in a metal bar. In addition, when one continuous cloud generates electric force within itself, and the others are at such a distance that they can't conduct lightning between them, the electrical indicator can show a great amount of electric force in the atmosphere without any thunder and lightning. All these phenomena can happen in a number of ways due to the different sizes, shapes, number, and positions of the clouds, and, therefore, those studies which seek a set of laws for the agreement of the indicator readings and the appearance of lightning seem to be fruitless. For that reason, I now come to the search for those very methods, so that it will be possible to ward off thunderbolts or to take refuge from them. It seems that both can be ensured by taking up a proper position or by setting up proper machines.

---

[23] *sagen'* – old Russian measure of length, about 2 meters
[24] reference to the G.Richmann's electrometer

As far as positioning is concerned, in mountainous places, according to the proposed theory, it seems that it is more dangerous to be in the shade for the air descending into the shade leads an electric cloud lower towards it and draws it down. Consequently, those places which, before the origination of thunderstorm clouds, were illuminated by the sun's rays and were heated up, can be considered as less dangerous than areas in the shade. But this can be better studied further by the gathering and analysis of thunder bolts according to the differences in their locations. Shadowy and light sides of high dwellings and temples as well as dark and cold forests are subject to these considerations. Underground tunnels similar to mine shafts seem to be the safest of all, because besides the fact that elevated places are more exposed to thunder bolts than are lower places, I also have never heard or read of an occasion when lightning struck in a mine. This is supported by an example that I found in the Freiberg chronicles[25]. In the middle of the night of 29 December 1556 a huge thunderstorm cloud arose which, by its lightning, struck and burned sixteen churches, however not one case of damaging of a mine can be recalled, although the mountains of that region are dug through with mines everywhere and on all sides. On his Japanese trip Kampfer[26] wrote that the emperor of that land shelters from arising thunderstorm clouds in underground paths with vaults on which are covered on top by a great deep pond. For the Japanese are of the opinion that heavenly fire cannot penetrate the water material. I consider that these sanctuaries, although they were devised taking neither present day knowledge nor theory into account, are still not useless, because water takes on thunderstorm electricity most easily of all. And if a thunderbolt strikes it which often happens) then, it spreads over it and into the whole earth globe and gets extinguished, not having done any damage.

I said about concealing oneself from thunder bolts; now about the methods to protect oneself from them of which two, apparently, could seemingly be successfully used. One consists of electric arrows set up and propped up in a necessary manner and the other in a violent commotion of the air. In the first case thunderstorm electricity is conducted into the earth and the second method leads to mix up and to weakening of the electric movement.

In considering the first of these possibilities, it is known to all that lightning strikes most often at the pointed summits of high towers, especially if they are adorned with iron vanes or are covered with metal. For dry wood or the porous rock out of which the summits are constructed are of such a nature that they are unable to take on as great an electrical force as is metal. For this reason, when an infinitely great electric force is generated in metals, the dry wood and the porous rock which lie under the metals can be considered as direct electric protection. Consequently, the sharp-pointed towers then are similar in all respects to the electric arrows which the investigators

---

[25] Reference to A.Meller, *Theatri Freibergnsis Chronici Pars Posterior*, Beschrebung der alten loblichen Bergk-Haupt-Standt Freyberg in Meissen. Ander Buch, Freybergk, 1653

[26] Reference to the book of German traveler E.Kampfer *Histoire Naturelle, Eivile at Ecclesiastique de l'Empire du Japon*, published in two volumes in French in 1729 in Hague and in German in 1740 in Rostok

of thunderbolts set up by design and whose action in the attraction of thunderbolts is rather well known through many dangerous experiments and by the death of Mr. Professor Richmann. I consider that it is a useful measure to set up such arrows in places somewhat distant from human paths, so that the striking lightning would dissipate its force more on them than on people's heads and on buildings.

Not only the idea but also the use of the second method has been intensified in certain places, that is breaking up the thunderstorm clouds by ringing bells. I will briefly indicate how much electric force in the atmosphere can be dissipated by this method. The fact that electric force consists in the movement of ether is reasonably confirmed by physicists. This movement is much impeded by the presence of air. That is clear from the fact that electric light does not show up in a thin glass sphere if the air is not drawn out of it.

As this occurs with calm air, then it is probable that much greater activity can result in the agitation of the ether from the great trembling of the air. For this reason, it seems that it is not useless to jolt the air during thunderstorms not only by the ringing of bells, but also by frequent cannonades, so that the great vibrations will lead to disruption of the electric force and will abate it.

Much remains to be said which comes to mind about the testing of these matters but the shortness of time does not permit me to propound everything. Therefore, having let alone the flashing and crackling of clouds, I want to the discuss subtle atmospheric phenomena to refresh you from so many inflations and fires by the remembrance of pleasant dew.

The nature of these atmospheric changes, although it is far from that of electric force, nevertheless, is the result of similar movements. For this reason, a short explanation is worthwhile here.

After sunset, the lower atmosphere cools more quickly than does the earth's surface, which is saturated with plants moisture. Therefore, the cold air coming in contact with the still warm earth, becomes heated, expands, becomes lighter and rises until it reaches an equilibrium state while cooling. We know from the works of the late Professor Richmann[27] that vapors are the more abundant, the greater the difference is between heat and cold in water and in air. Therefore, air which is cooled after sunset gathers a great amount of moisture from the warm earth and carries it aloft to a certain height. Another type of dew which is squeezed out through chinks located in the grass does not belong to the category, and thus, having passed over that type, I must come to the remaining atmospheric electric phenomena.

It was shown above that during the winter time it often happens that with the sinking of the upper atmosphere a sudden frost is brought on without any severe blowing of the wind after warm weather. The phenomena of the northern lights occur to a great extent in winter after a thaw, so that they very often foretell a frost or come along suddenly with one. Electric friction of vapors occurs in the atmosphere with the sinking of the upper and rising of the lower atmosphere, which is known from the above-mentioned theory about the origination of lightning and thunder. Thus., it is very probable that the northern lights originate from the atmospheric electricity. This is confirmed by the similarity in the appearance and disappearance, in the movement, color and forms which are exhibited both in the northern lights and in electric light of the third type. Electricity generated in a sphere from which the air has been withdrawn emits sudden streaks of light which disappear in the twinkling of an eye while almost at the same time new ones leap out in their place, so that there seems to be a continuous flashing. In the aurora borealis the northern lights or the beams, although they do not appear so unexpectedly fast due to the extent of the whole aureole, nevertheless have a similar form. for the flashing columns of the northern lights extend almost perpendicularly in bands from the surface of the electrified atmosphere into the thinnest or very rare ether; the converging beams do flash in a similar manner than they do in the mentioned electric sphere from the concave round surface to the center. In both phenomena the color is pale. All the exhibited forms of the northern lights cannot be vapors or clouds illuminated by some kind of flashing, as their almost always regular form and the stars shining through distinctly show.

---

[27] Reference to dissertation of G.Richmann *De Evaporatone Ex Aqua Frigidiori Aere Observations et Consectaria* published in Vol. II of *Novi Commentarii Academiae Scientiarium Petropolitanae* (1751, pp. 145-161). The same issue was addressed in Richman's oratory *De Legibus Evaporationis Equae* read on November 29, 1749 and published in the proceedings *Celebration of the Academy of Sciences… and Public Meeting … on 29 of November 1751* (St.Petersburg, 1749, pp.1-34).

Great plausibility is added to this from my observations, by which it appeared that at the beginning of autumn and at the end of summer, heavy with frequent thunderstorm clouds, the northern lights appear more often than in other summers. In addition, sometimes even during the time for the northern lights themselves, I noticed the flash of heat lightning. It appears from this that the aurora borealis and the northern lights of heat lightning can be distinguished not by their nature but by the degree of their power and location. Heat lightning follows after a strong electric force, during its disappearance, at night, and in thin atmosphere. The northern lights appear in the middle atmosphere above its boundaries due to the weak friction of vapors. By experimentation we are sure of the fact that a visible radiance can be produced in a place lacking air, and for this reason we can pass up here without fault all the considerations which demand a clear and detailed knowledge of ether. The location of the northern lights above the limits of the atmosphere is concluded from comparison with heat lightning. For that periphery must be equal to the great circle on the earth's surface can be concluded from the nature of the earth's shadow; the encirclement of the northern lights has to be equal to the circles which are parallel to the equator and the width which the encirclement has at the surface of the atmosphere, which can be deducted from ratio of the height of the regular arc of the northern lights to its width.

This is also confirmed by observations which were made last winter. On the twelfth of February, as the twilight was drawing to a close, bright northern lights appeared, quickly spread over the whole sky, and a bright arc was reflected not only in the north, but also on the southern side; however, the electric indicator that was set up and which showed the thunderstorm force in the summer did not give any sign even of a little electrification.

Therefore, the electric force which creates the northern lights is exited near the upper part of the middle atmosphere, the air of the uppermost layer will move and expand the columns and arrows with a shaking of the rare ether. All the air in the atmosphere which is of that density which extinguishes electric sparkling in a glass sphere[28] remains dark, surrounded by a bright arc which presents an easy method of determining the height and distance of the polar aurora.

Once we have assumed this, the reason for several of the common phenomena remains to be pointed out. For an interpretation of all the phenomena which consist of many different forms and movements would require a lot of time.

In the first place, you may ask why do the lands which lie to the north experience the aurora more than those lands which are closer to the equator? To answer this, I must first point out that the descent of the upper atmosphere into the middle atmosphere should take place more easily near the poles than near the equator. It is clear from what has been written above that the cold layer of air near the polar circles joins the ocean surface, thus it follows that the upper limit of this layer, which is also the lower limit of the upper atmosphere, passes close to the earth. Then the air of the uppermost atmosphere does not feel the action of the sun's heat much anywhere, which is shown by a comparison of the barometer and the thermometer, but near the polar circles and toward the poles in autumn and winter the power of the sun's rays is felt even less due to their great declivity and to the shortness of the day and the perpetual absence of the sun's rays. Therefore, it is very probable that the air which comprises the upper atmosphere is compressed by the great cold in those parts to a density which the middle atmospheric layer snow has. Because of its density, the vapors can rise to the upper surface of the atmosphere. Therefore, when the earth's heat communicating by means of the open seas to the air lying above it heats this air and expands it so much that it must have proportionally less weight than the upper air, then at the same time the upper air mixes with the lower, which rises to meet the upper, an electric force is generated which extends to the upper limit of the atmosphere, and an aurora is created in the free ether.

In our lands, the northern lights, for the most part, show up after twilight and rarely continue through the whole night. The reason for this circumstance can be quickly seen. For the lower atmosphere, which is heated up by the sun's rays during the day, is thinner after the sun sets than it is later on in the night when it cools down more and becomes denser due to the absence of the daytime heat and due to the sinking of the upper atmosphere; the friction and electric force cease, and the light extinguishes. But if the cause gets forceful, i.e., if the difference in densities of the upper and lower atmosphere is greater, it is indisputable that the lights can continue throughout the whole night.

Thus, the continuation of the disturbed equilibrium in the atmosphere, especially at the polar circles, brings about continual northern lights so that there is enough light for those people living around the North ocean during the winter time when the sun does not shine and also at the time of the new moon for getting things done. For when the upper atmosphere feels little or nothing of the sun's ray and is compressed by great cold, then the lower atmosphere, lying on the open sea, is heated up, expands, rises, and the upper atmosphere sinks. And in as much as the severity of the cold in the upper atmosphere and the thaw in the lower atmosphere continues uninterruptedly, for this reason it is not surprising that the electric friction does not stop and the aurora is always visible.

---

[28] Reference to the phenomenon which is described now by the Paschen law [10]

Having passed over explanations of other phenomena, I can not be silent about one of them, i.e., the phenomenon of the different colors which sometimes inflame the whole sky during the northern lights, to the fright of those looking on. A brilliance of this type occurred in the north and in the south on 23 January 1750, and I observed it studiously. The changes took place in the following order. At the expiration of the sixth hour in the afternoon and after twilight, an orderly aurora appeared very clearly in the north right away. A white arc shone over the dark opening, above which, behind the blue band of the sky, another arc appeared with the same center as the lower arc, scarlet-colored and very clear. A column of the same color rose up from the horizon, to the south west and extended itself close to the zenith. Meanwhile the whole sky was lit up by bright bands. But when I had glanced to the south, I saw a similar arc on the opposite side of the north, with the difference that rose-colored columns, which at first were numerous in the east and later in the west, rose up on the scarlet colored upper band. A short time after this, between the white and scarlet-colored arc of the aurora australis, the sky was covered with a green color similar to that of grass and the form taken resembled a rainbow, after which the scarlet-colored columns disappeared little by little, the arcs still shone, and not far from the zenith a white aurora of the size of the sun emitting diverging rays, and columns rose from the south-west to the zenith and practically touched it. Thereupon a scarlet-colored spot appeared to the west between the beams of that aureole. Meanwhile eight hours had passed and the sky glowed with scarlet-colored and gray-colored of irregular shape; there was more of the grassy color than the scarlet color. At the zenith, in place of the aurora emitting the beams, two arcs appeared which intersected each other. The arc which stood with its concave side to the north had transverse rays of light bending towards the center and the other arc which had its concave side turned towards the south had longitudinal rays of light parallel to the periphery. The ends of both the arcs stood out about five degrees from the intersection and from the zenith. All these changes ended at nine o'clock, and one regular aurora was left in the north, of the type which often occurred here.

I consider it well to avoid explanations of all these forms, which I will try to explain in the future by the outlined theory. And therefore, I mention the colors only very briefly. In considering the arcs, which are similar to a rainbow, it would be convenient if I attribute the colors of the night aurora as coming from the refraction of the rays, but that does not stand three circumstances. In the first place, there was not such a star, the refracted rays of which could have been separated into colors. Mixed aurora of columns and arrows can not be the cause of such a regular phenomenon. In the second place, scarlet-colored columns were of the same shape and moving in the same direction as the white ones; therefore, they come from the same source, which is very different from what would come from the refraction of rays. Thirdly, which fact has not been proven anywhere, that all the colors come from the refraction of the rays, but on the contrary, there is a good deal of proof which clearly shows that colored bodies show different colors to the sight, only by avoidance of rays. In a similar way no one thinks that these night colors were merely illuminated vapors or clouds, no one who considers their form, which is different from the characteristic form of vapors and clouds and their position outside the atmosphere.

Thus, the fact remains, that their causes must be looked for in the differences in the ether. A difference in colors in different types of ether or at least at different rates of its movement will be assumed, everywhere the possibility will be found for the ether to exhibit different colors, i.e., by the movement of red ether (or, according to another opinion, the flickering speed producing the red color) it can produce a red color, and by the movement of yellow ether with blue, a green color can be produced. And in a word, when the most complicated of all the main colors, i.e., the white color, is generated in ether without air, then one must by no means doubt that the color component can show up separately. The electric aurora produced by experiment agrees in no small way with this, the aurora sparkling with different colors according to the differences in the bodies, from whence it can be concluded not without probability that different colored columns and aurora are generated in ether at the very surface of the atmosphere by the movement of different vapors.

Having explained these phenomena in so far as is possible from the laws of electricity, the phenomena which show us the actions of the earth's atmosphere, I wish to proceed further now to the examination of those bodies which, swimming around in the vast ocean of ether, exhibit similar forms.

Comets are considered in the first place, which the sage philosophers no longer doubt to consider as important bodies of the whole universe along with our globe and the other planets, but the cause of the pale brilliance and their tails is still not sufficiently studied, a cause which I assume, without doubt, can be found in electric force. It is true that the reasoning of the ingenious Newton is contrary to this[29], Newton who considered tails as vapors coming from the cornets and illuminated by the sun's rays; however, if during his time, as much light had been shed on the subject of physics from the discovery of electric force as there has been today, then I expect that he would have had the same opinion as I am trying to prove today. I have already noticed for several years that attempts to explain the origination of the tails of comets as coming from vapors is a task subject to great and, apparently, impenetrable difficulties. For this reason I considered it worth to let this idea alone completely and to search for another cause, always having in mind the suspicion that this phenomenon is related to the northern lights, and that both of them stem from the movement of ether. The thoughts which I have had for a long time about the sinking of the upper atmosphere into the lower, thoughts which today are illuminated by the rise of the electric era in natural science, led to the following theory concerning the tails of comets.

It is impossible to measure the atmosphere of a comet either by the length of the tail or by the width of the aurora which surrounds the head as is indicated in the following discussion, however there is no doubt that it exceeds the height our atmosphere many times. Similarly, it is clear that its density increases much more according to the measure of height and pressure and the vapors rise higher. When a comet approaches close to the sun and is reached by the heat of the sun, then a part of its atmosphere located in the shade of the body does not feel the direct solar rays. Those rays which are diverted due to the vast expanse of air shine like a great afterglow into the shadow of the comet and can not be the cause of any heat at all. Therefore, on the side away from the sun, a dark air column extends from the surface of the body to the surface of the atmosphere itself, having the width of the whole shadow. The air which composes that column must be much colder, thinner, and proportionally heavier than that which is outside the shadow in the remaining atmosphere and exposed to the direct sun rays. Having considered the great height of the atmosphere which can be assumed without fear of error to be ten times higher than our atmosphere, it can be clearly understood that it greatly outweighs the other parts of the atmosphere and that it must sink down with a rapid movement towards the body of the comet. Meanwhile, it is necessary for the light air expanded by the solar rays to go toward the column and to flow in to occupy the place which is left in the shadow due to the sinking column, where, having cooled and become denser, it must become heavier and similarly sink below after the other. and then it is obliged to vacate the place for the air following. Thus, by the continual and rapid flow of air directed upwards and downwards, a forceful impact and friction of the vapors is excited around the boundaries of the air column circulating in the shadow, and a great electric force is generated. The pure ether outside the atmosphere produces light by rapid flickering, corresponding to the air movements, i.e., extending in space on the side opposite to the sun, behind the comet in its shadow. Thus, tails of the comets appear in different forms according to the difference in the atmosphere of each comet and to the differences in distance and position with respect to the sun. The air column in the shadow of the body of the comet composes a large part of its atmosphere, in as much as it has half of the surface of the whole body as a base, for this reason, the whole atmosphere and the great quantity of vapors which surround the come on all sides must be subjected to no slight vibration caused by the very violent movements of the current. From whence electrical frictions arise which, although they are much more calm than those indicated above, however, are not completely unsuited to the electric movement of the ether. Thus I consider that the whole aurora that surrounds the head of a comet can not be considered as vapors illuminated by solar rays, and particularly, I consider that a great part of this aurora is very similar to the tail itself.

---

[29] See in I.Newton, *Philosophiae Naturalis Principia Mathematica*, Londini, 1687 (2nd edition, Cantabriae, 1713)

Today everyone can see that the tails of comets are now considered as one with the northern lights which occur in our lands, and that they differ only by one magnitude. It is true that, in an addition to the proof of the proposed theory, these two phenomena have surprising resemblances in the most notable circumstances, so that their agreement can serve in the place of strong arguments. For as far as position is concerned, both show themselves on the side turned away from the sun. The stretched-out strands in the tail of a comet completely coincide with the columns and the rays with which the northern lights exhibit. Finally, the paleness of both the phenomena, yielding to the rays of the stars, attests to the similar nature of both these phenomena. In both cases the weak electric flashing is overcome by the great brilliance of the stars.

Therefore, since the tails of comets are not vapors rising from them, but only the movement of the ether produced by the electric force, for this reason those fears which arise during the appearance of comets are unjustified, since many think that great floods are caused on the earth by these comets.

There are still many phenomena similar to this, such as ·the zodiacal light, the milky way, and many dull stars, the cause of which, apparently, is not different from the origination of the northern lights and the tails of comets, but the enormity of the material, which has fatigued me, forces me to stop the course of my discussion, and perhaps you. who have been. listening so long are desirous that I stop.

Thus, in winding up my speech, I turn to the one who created man, so that he, in considering the great extent of the things created, the uncountable quantity, the endless difference and the chain of the connections placed between them by the supreme Providence, should marvel at his wisdom, strength and grace with reverence. I invoke supplication to him with great fervor, so that, upon the opening and discovery of the many natural secrets with which he most generously blessed our lives, even for future time, by the unceasing efforts of scientific people who are everywhere becoming learned as to the creations of his hands, he should condescend to contribute by fortuitous successes, let him open up a safe sanctuary for the preservation of health and life of mortals from harmful atmospheric impetuosities, so that through his assistance to the divine projects of Peter the Great and to the maternal generosities of his most august daughter we can respond with the fruits of our labors; so that under the tranquil command of Elizabeth the sciences growing in our beloved homeland will grow to full maturity and will produce a very rich harvest; let there be an equal happiness for him and equal rejoicing for us as there was for this city and for its inhabitants in the fifty years which followed the founding of this city which are ending today. And as this city was founded by Peter's blessed enterprise, and as it grew in such a short time to great size and attained a flowering state, in a similar fashion let this Academy set up by the same great founder under the protection of his true heiress expand and flourish for his immortal fame and for the good of the homeland and for all mankind.

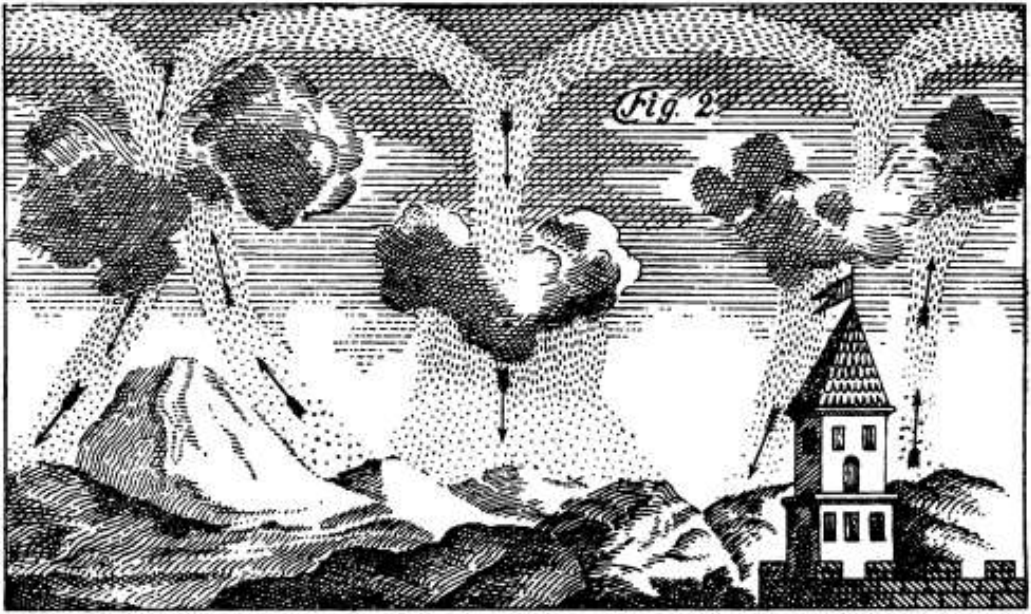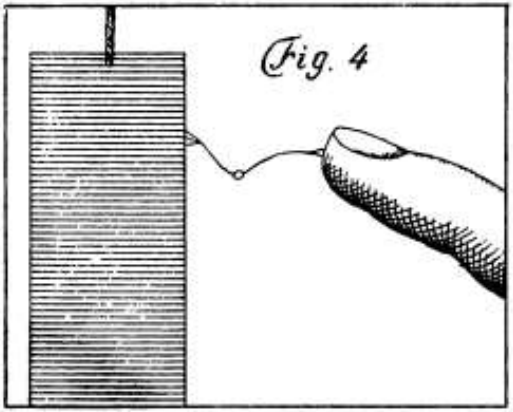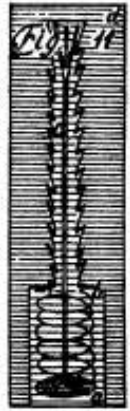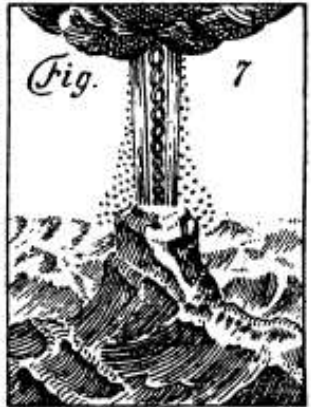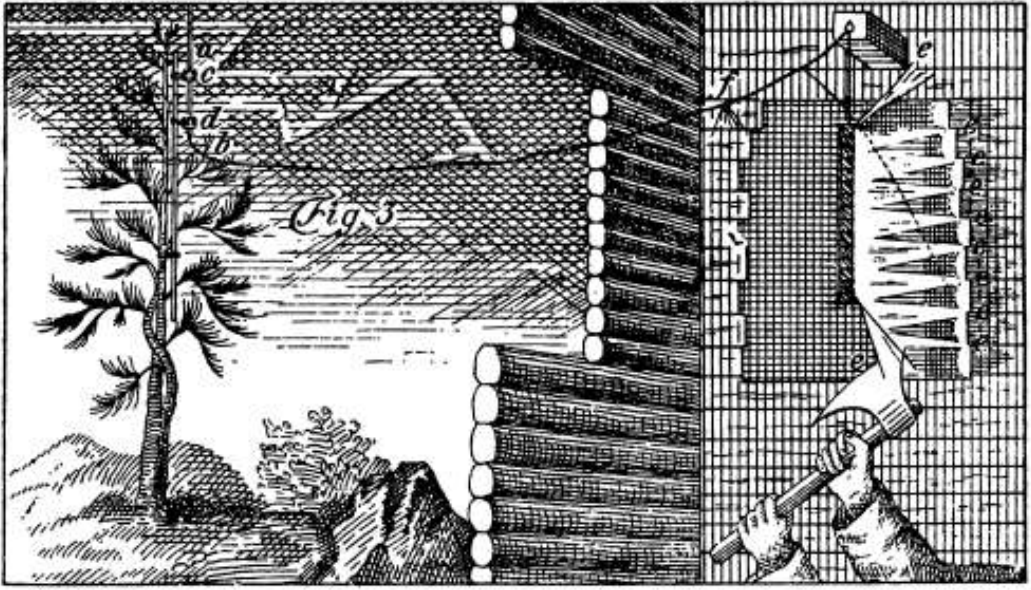

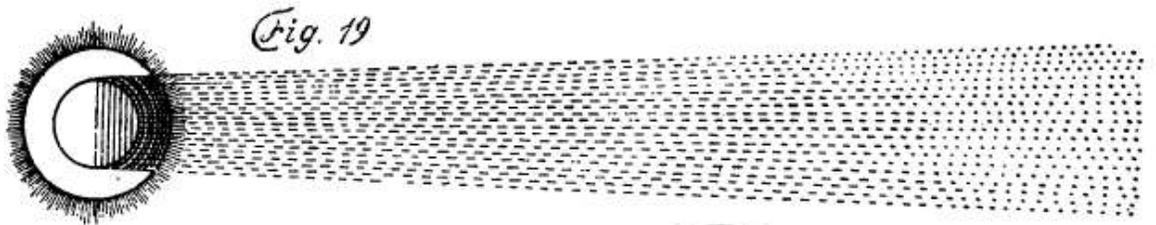
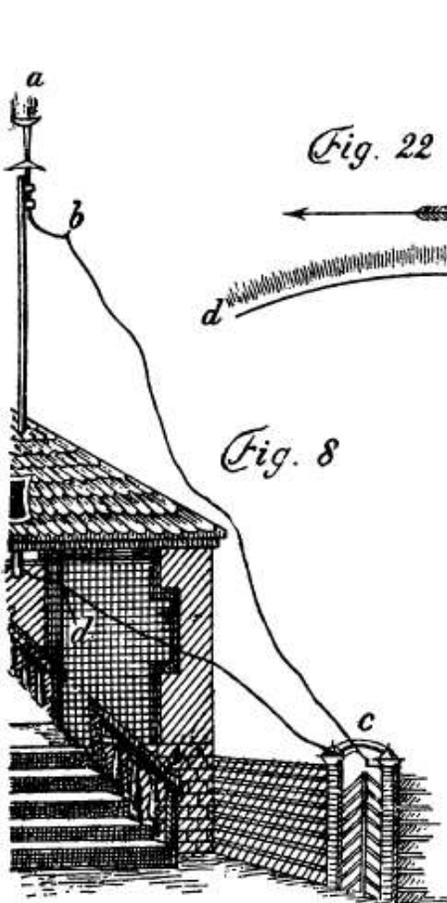
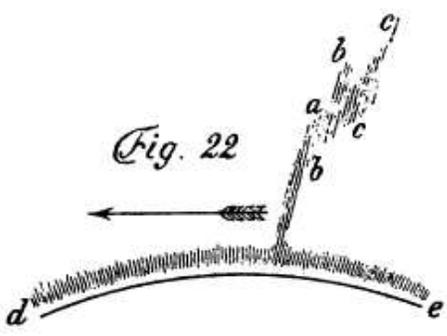
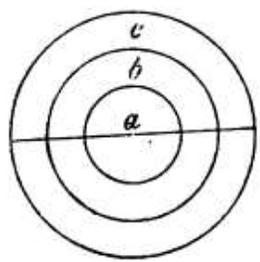
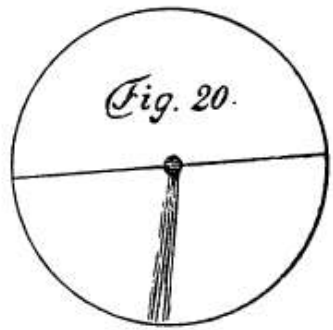
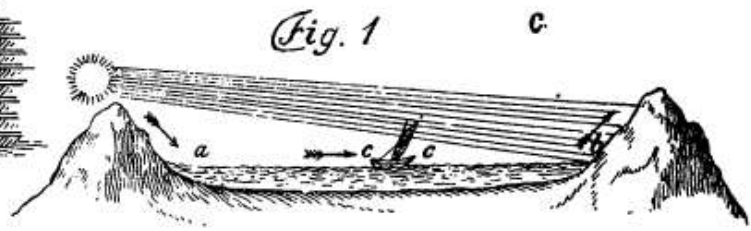
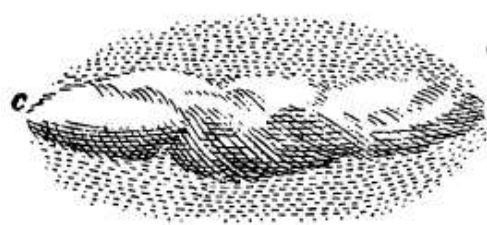
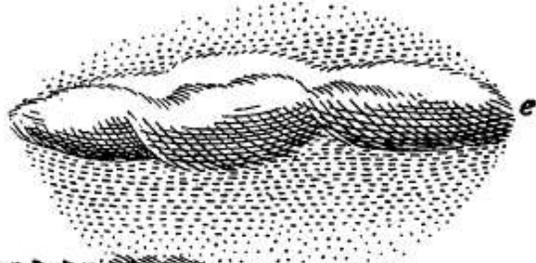
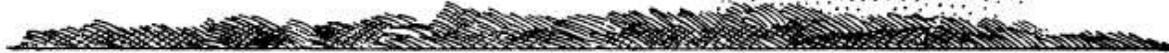

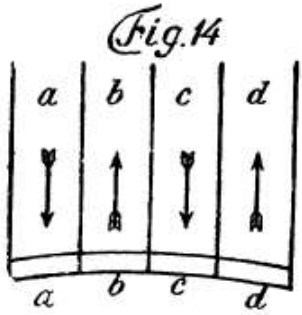
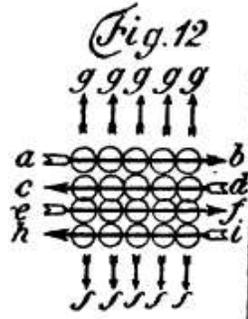
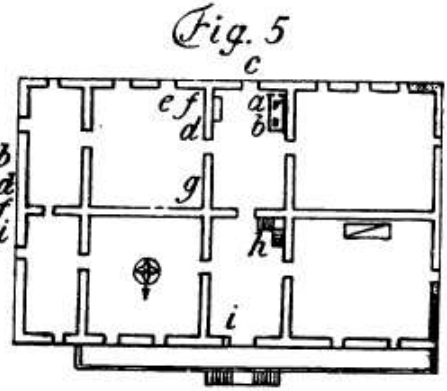
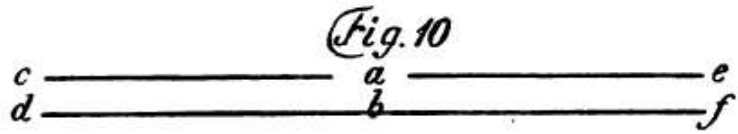
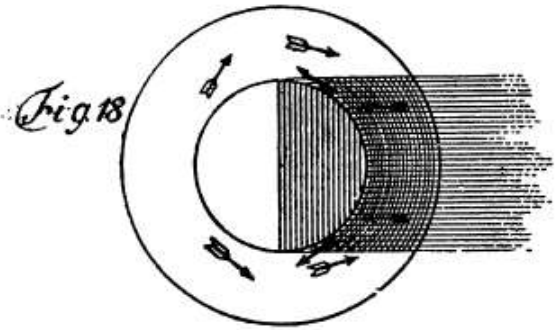
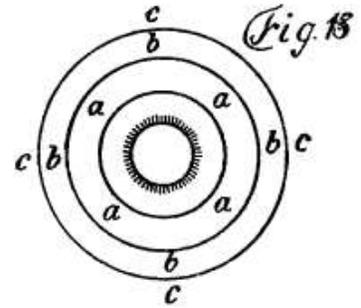
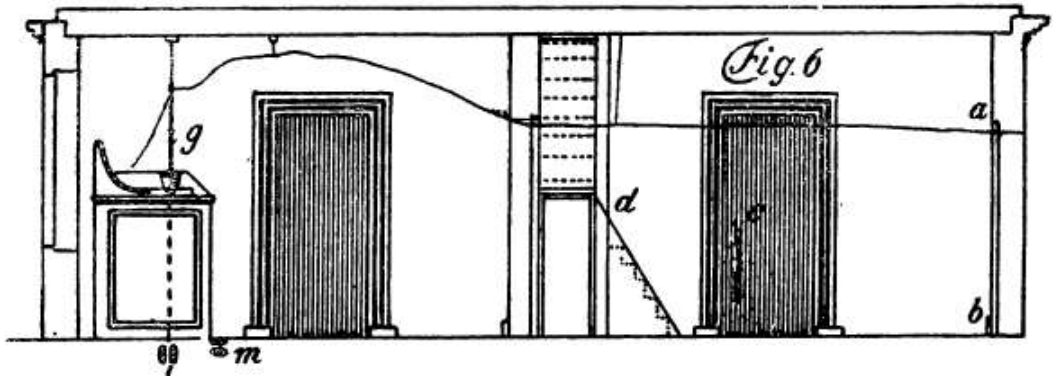

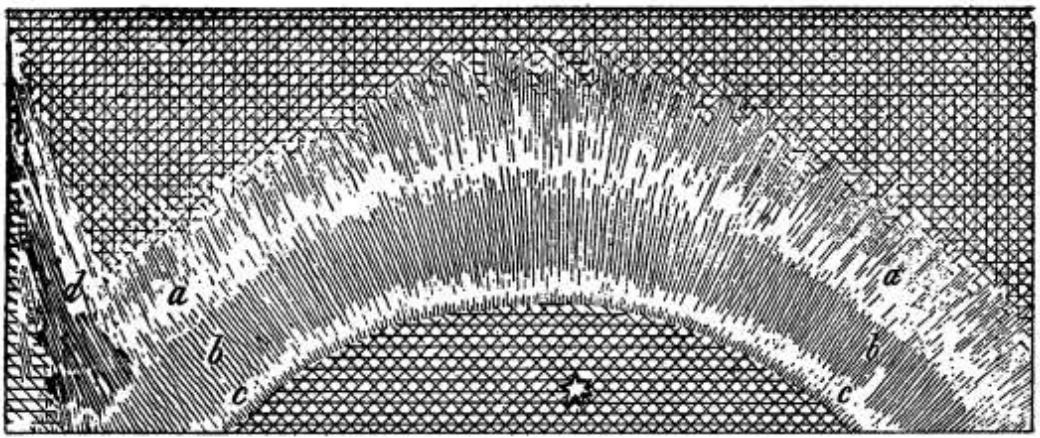

Fig. 15

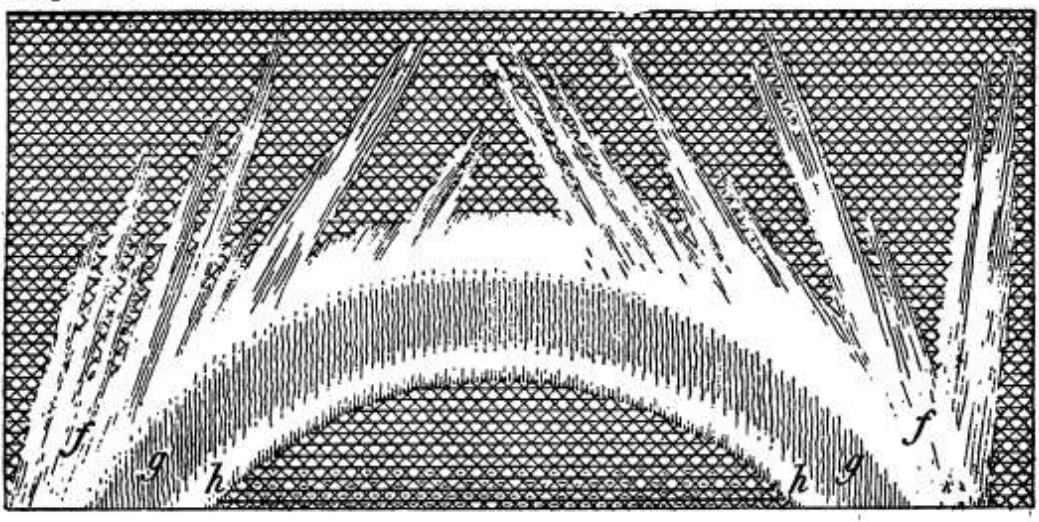

Fig. 16

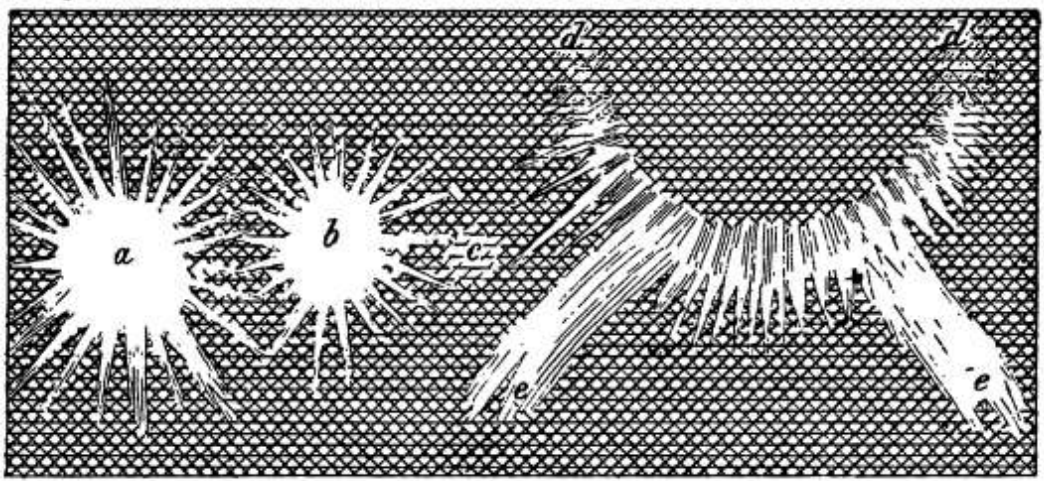

Fig. 17

## *Explanations of the "Discourse…" illustrations on pages 26-29*

Excerpted from the M.Lomonosov's "Explanations…", [8], vol.III, pp.101-133, in order they appear on the original engraving plates (top left to bottom right):

*Fig.2:* Lomonosov's explanation of the atmospheric electricity generation by friction of the air-borne "organic" particles in the colliding up- and down-drafts of air. Hot air from the surface gets up, while cooled air from above gets down. Effects of temperature differences on sunny and shady parts of the mountains and big buildings are illustrated – e.g., warmer air updarfts take place on the sunny sides.

*Fig.4:* The shape of an electric spark between a finger and a charged metal rod.

*Fig.11:* Proposed electrical meter for measurements of large instantaneous quantities of electricity – metallic disc at the bottom is attracted to the metallic base when charged and its maximum movement can be read by the displacement of a trigger-like shaft in the ragged structure. The shaft and the disk are weighted by a spring.

*Fig.7:* Typhon (typhoon) in a sea – presumed to involve electric attraction of the water below.

*Fig.3:* Experimental set-up at the Lomonosov country-side estate (Ust'-Ruditsa): metallic rod *a* was connected to a metal wire *bf* coming inside a wooden house. Under a thunder cloud, sparks and hissing were coming out of the wire and out of the sharp edges of dry wood longs *eeee* when an axe approached them.

*Fig.19:* a comet with tail and surrounding atmosphere and radiance beyond the atmosphere; see also *Fig.18*.

*Fig.22:* "The tails of the comets bend while approaching the Sun, when moving sideways. The pillars of the Northern lights, with such a movement, leave behind some parts of the vanishing pillars, and altogether look like a curved comet tail. The pillar *a* moves along the arrow; the disappearing pillars are marked *bb* and *cc*."

*Fig.21:* See caption *Fig.20*

*Fig.8:* Lomonosov's own set-up at his city house: *a* is metal rod with many sharp needles; *b* - isolated wire; *d* – the room for observations.

*Fig.20:* Comet's tail, extending to at least half-diameter of the comet's atmosphere: illustration of presumably wrong supposition that comet tails can take place inside atmosphere of the comet. If that would be the case, argues Lomonosov, Sun rays would go through and equally illuminate the entire atmosphere of the comet – see *Fig.21* - and the comet would be seen as an enormous glowing sphere. As that is not the case, then the tails are located mostly outside the comet's atmosphere, as depicted in *Figs.18* and *19*.

*Fig.1:* Illustration of the air movements over one of the lakes in Waldstatt, Swizerland, reported by J.J.Scheuchzers in *Natur-Geschichte des Schweitzerlandes, samt seinen Reisen iiber die*

*schweitzerische Gebiirge*, Zurich, 1746; the winds changed their director from West to East depending on which side of the mountain coast is warmed by the Sun.

*Fig.9:* On horizontal electric discharge between two clouds, one neutral (*ca*) and another charged (*ae*).

*Fig.14:* Vertical motion of ether particles as origin of the Northern lights, intermittent rows *aa, bb,* etc move in opposite directions; see also *Fig.12*. The bottom part of each row represents Earth's atmosphere in which electricity is generated through vertical air drafts. The upper parts indicate the areas above the atmosphere, i.e., free of air, where the electricity-induced excitation of ether particles results in the lightning.

*Fig.12*: Light can be produced by motion of ether particles: during their oscillations, intermittent rows *ab, cd,* etc should move in opposite directions.

*Fig.5*: Plan of G.Richmann's house, see also *Fig.6* for placement of various wires, indicators, etc

*Fig.10*: simplified illustration of the effect of horizontal discharge between two clouds, depicted in *Fig.9* – lightning and thunder will occur at the same moment for an observer in point *b*, and at very different times for observers in *f* and *d*; appearance of stronger electric force will be detected in d as after the discharge the electricity will be equally shared between the clouds *ca* and *ae*.

*Fig.18*: Air flows in the atmosphere of a comet. "…Although some glorious scientists noted similarities between comet tails and the Northern lights, but no one except me figured out that : 1) an electric force is born via collisions and frictions of ascending and immersing air flows in the shadow of the comet in its atmosphere; 2) that the electrical power originated in the shadow of a comet produces luminous movements in the ether; 3) that the tail and part of the radiance, surrounding the head of he comet, occur and are visible in places which are completely without the air and vapors, and therefore, the radiance has nothing to do with sun rays."

*Fig.13*: Waves of light oscillations *aaaa, bbbb, cccc* of the ether particles spread out in all directions from the Sun.

*Fig.6*: explanation of the experimental set-up in the G.Richmann's house, *g* is for the "electrical index" which Richmann stand next to; *m* is the location of another observer, engraver Sokolov; see also Fig.5

*Fig.15*: Colors in the observed Aurora borealis : *aa* scarlet arc; *bb* – the sky; *cc* - white arc; *d* - scarlet column.

*Fig.16*: Colors in the observed Aurora australis : *hh* – bright arc; *gg* – green arc; *hh* - white arc; *ff* -scarlet arc.

*Fig.17*: White shining in zenith: *a* – white shining; *b* – with scarlet spot in *c*; *dd, ee* - arcs in zenith.

*Commentary* [V.Shiltsev]:

This is the first complete English translation of the Mikhail Lomonosov's seminal work, from its Russian version [1]. It continues the series of English translations of Lomonosov's scientific works, started by H.Leicester [11] and followed by [12, 13]. More on the life and works of the outstanding Russian polymath and one of the giants of the European Enlightenment can be found in books [5, 14] and recent articles [15-19].

The history of this work of Lomonosov, written in May-October 1753, is as follows. On May 7 (o.s.), 1753, it was announced at the conference of the St.Petersburg Academy of Sciences Conference that the annual public meeting of the Academy is scheduled for September 6. All academicians were invited to "think in advance about dissertations, which will be read at this public meeting". M.Lomonosov and G.Richmann immediately indicated their interest and it was decided at the at the next conference, held on May 14, 1753, that Richman will give a talk about his observations of electricity and Lomonosov would augment that with discussion on the cause of electricity and its benefits to human lives. That was further specified in the order of the Academy's President, Count K. Razumovsky on June 14, 1753, to Richmann to read his thesis to public in Latin and to Lomonosov to prepare his response in Latin and Russian. All the materials were expected to be published prior to the September 6$^{th}$ meeting.

During May-July 1753, Lomonosov and Richmann were preparing to their presentation and in parallel to writing of the texts, continued their dangerous experimental studies of lightning, beyond what had been already accumulated by them during observations in the summer of 1752. To come to more convincing conclusions, a number of new studies with an advanced electrical measuring instrument "electrical index" (indicator or "gnomon" – see Fig.1a) below and a "thunder machine" had been carried out. Another reason of these extensive studies was that they followed the first news on Franklin's discoveries which reached Russia and published in the *St. Petersburg Vedomosti* newspaper on June 12, 1752. Both scientists set up labs at their homes, not far from the Academy of Sciences – see Fig.2 – and both seemingly were very well aware of the dangers of their experiments. Lomonosov also conducted some of the experiments at his estate in the Ust-Ruditsa factory, some 50 miles West of St. Petersburg.

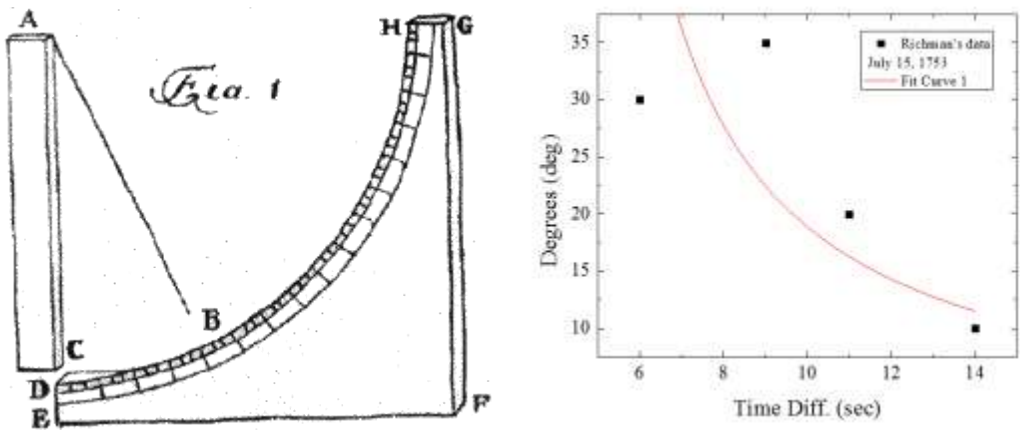

Fig.1: a) left – Richmann's electrometer ("electrical index"), 1745 [8]; b) right – Richmann's data, indicating reduction of the atmospheric electricity (deviation of the "index's" thread from vertical, in degrees) with the distance to the source (time difference between the lightning and the thunder), data take from the report published in *Vedomosti* on July 15, 1753.

G.Richmann regularly reported his results in the *Vedomosti* – on May 7 (No. 37), on May 11 (No. 38), on May 18 (No. 40) and on July 13 (No. 56), 1753 – see Fig.1b. His communicated on efficiency of lightning protection indicate that sharp-end metal rods work best, and that dielectrics (glass) are good to hold the metal rod. Lomonosov also reported in the newspaper of June 4 (No. 45) of 1753 and was the first to establish that "power of the electricity in the air may extend beyond the area of thundering or be present even without a thunder" - that is, to detect an electric field in the atmosphere. In April 1753, the two studied whether canon shorts could affect the atmospheric electricity during the Imperial Court's celebration fireworks which employed up to 58 canons – and seemingly found no significant effect.

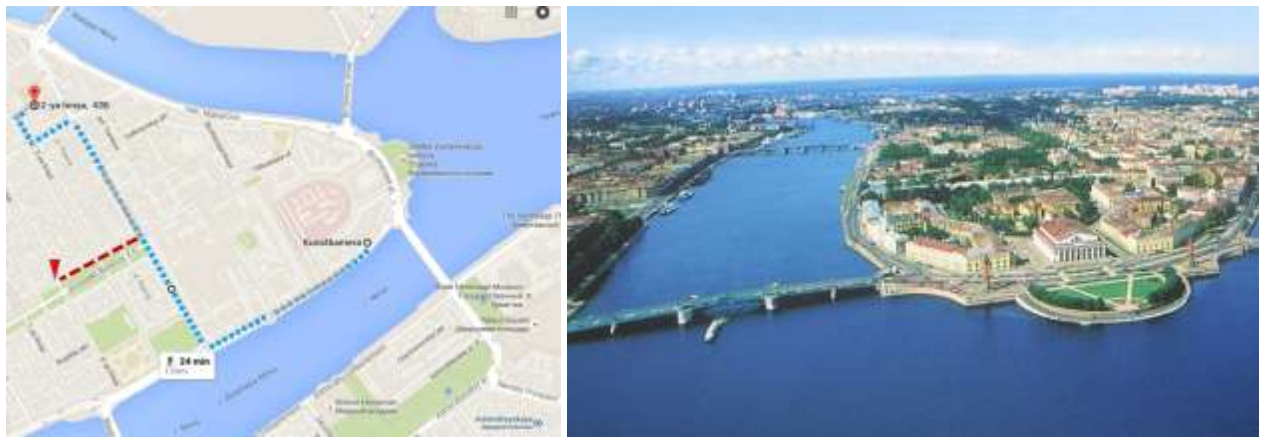

Fig. 2: The map (left) and photo of the Vasilievsky Island in St.Petersburg, indicating the paths of Lomonosov and Richmann from the Academy (Kunstkamera) to their homes on the 2$^{nd}$ Liniya and on the 5$^{th}$ Liniya, respectively, on the infamous afternoon of July 26, 2753.

On July 26 (o.s.), 1753, Academician Georg Richmann was tragically killed by a lightning strike while conducting the experiments – see, e.g., [3, 4] for more details. On that day, from 10am to about noon, both Richmann and Lomonosov attended Academy's meeting at what is now Kunstkamera. Around noon they have noted a big thunderstorm cloud coming and left for their home labs: Richmann to the 5th Liniya (about 16 min from the Academy) and Lomonosov to the 2nd Liniya (24 min). Both observed strong electricity activity out of the cloud in their setups, both experienced minor shocks prior to the disaster, Lomonosov was eventually distracted by his wife (presumably, asking him to take long awaiting lunch).

Richmann's death made a great impression in the academic world as in St. Petersburg, and abroad. Detailed reports about the accident were published in the *Vedomosti* on August 3, 1753, as well as in some foreign periodicals, for example, in the *Philosophical Transactions* [20, 21], in the *Memoires de l'Academie Royale de Sciences* (Paris) and others. The description of the tragic death of his friend was given by Lomonosov in a well-known letter to his patron, Count Ivan Shuvalov ([8], vol.X, pp.484-485) written on the same day while still being under the impression of the tragedy: "…What I am now writing to your Excellency – needs to be considered as a miracle, for the fact that the deads do not write. I do not know yet, or at least I doubt whether I am alive or dead. I see that Mr. Richmann was killed by thunder under the exact same circumstances in which I was at the same time…. Meanwhile, Mr. Richmann died a beautiful death, performing the duties of his profession. His memory will never get silent." In the time followed, Lomonosov did a lot to arrange pension for Richmann's widow and children and effectively cared for their well-being.

Shortly thereafter, on August 5, 1753, an Adviser to the Chancellery of the Academia I.Schumacher wrote to the President, who was then in Moscow, on the desirability, in his opinion, to cancel the September 6 public meeting of the Academy. Accordingly, Razumovsky canceled the meeting on September 2, i.e., just four days prior to it. Lomonosov, naturally, could not reconcile himself with the cancellation of the public meeting and pulled all possible strings and insistently sought revision of the President's decision. After much trouble, Lomonosov managed to succeed and on October 18 , the Conference of the Academy had announced a new presidential decree on the organization of the public meeting on the 25$^{th}$ of November with following motivation ".... so that Mr. Lomonosov would not be late with his new inventions among the scientists in Europe and by that his work on electric experiments up to this time would not in vane". Acdemician A. Grischow was appointed an official opponent of Lomonosov presentation at a public meeting.

Lomonosov first wrote the Latin version of the speech, sent it to his fellow Academicians and then himself translated it into Russian. Members of the Academic Conference A.Grischow, N.Popov and I.Braun submitted their doubts and objections on individual particular moments of Lomonosov's speech – all of which Lomonosov addressed to the Conference's satisfaction and the approval of the publication was given on November 3. On November 16, the Conference also approved the text of the Grischow's reply to Lomonosov's speech. It was decided that both public presentations will be given in Russian. Finally, on November 26 (o.s.), 1753, Lomonosov read *Discourse on Atmospheric Phenomena Originating from Electrical Force* at the public meeting of the Academy of Sciences. Lomonosov's presentation, Grischow's response were published separately in Russian and separately in Latin. Later, an addendum to the Lomonosov's work, entitled *Explanations, Appropriate to the Discourse on Atmospheric Phenomena Originating from Electrical Force* was also published ([8], vol.III, pp.101-133). It should be considered as an integral part of the Lomonosov's work as it contains descriptions of a number of new observations and experiments, executed by Lomonosov, and explanations of the figures and drawings attached to the *Discourse* – see Fig.1 above. In the *Explanations* Lomonosov also proves as unfounded the doubts Grischow, who tried to belittle originality of the Lomonosov's research in the field of atmospheric electricity and to attribute him the role of imitator of B.Franklin. Lomonosov points out that a) his "Discourse.." had been written and sent for publication before any communications of the Franklin theories reached Russia; b) his "theory about the cause of the electrical force in the air" has nothing taken from Franklin, even in the cases which look similar – like the origin of the Northern lights – their explanations are totally different, as Lomonosov had totally different approach based on the air up- and down-drafts; c) Lomonosov's theory got initiated after observations of electrical phenomena right after major cold air downdrafts causing severe frosts – nothing that Franklin ever could observe in Philadelphia, d) Lomonosov has evaluated mathematically the phenomena of the up- and down-drafts in atmosphere;  e)  he interpreted many phenomena which Franklin did not even considered.

200 out of 300 copies of the Latin version of the *Oratio De Meteoris Vi Electrica Ortis* were sent abroad, to foreign honorary members of the Academy, universities, foreign academies and large libraries. In January-February of 1754 several responses to the Lomonosov work were received from L.Euler, G. Krafft, G. Heinsius. Euler's comments were rather positive: "… the mechanism proposed by the wittiest Lomonosov concerning the currents of that subtle matter in the clouds, should bring the greatest help to those who want to study the issue. His thoughts about lowering the upper air and about the sudden cruel frosts happening from this are excellent." Some engravings from the Lomonosov's paper were reprinted by William Watson in the account of G.Richmann's death in the *Philosophical Transactions* [20].

Lomonosov himself very much valued the *Oratio De Meteoris Vi Electrica Ortis* and listed it among his most important scientific accomplishments – see [8], v.10, p.398 and p.409 – as well as included it in the convolute of his 9 major publications, bound under the title *Opera Academica* just in twelve copies, which were sent abroad on very special occasions, such as, e.g., for consideration for election to the Bologna Academy of Sciences in 1764 [22]. The interest in the atmospheric electricity led Lomonosov to create the first model helicopter. In 1754, looking for a way to send meteorological instruments and electrometers aloft, he designed and built the first working helicopter model. It used two propellers rotating in opposite directions for torque compensation, and was powered by a clock spring. While Leonardo da Vinci famously left a sketch of an airscrew, Lomonosov actually constructed a proof of principle that managed to demonstrate significant measurable lift – see, e.g. [15].

Most, though not all, elements of the Lomonosov's work are profoundly correct even by present day understanding of the phenomena. The basis of the Lomonosov theory is the idea of vertical air movements as the main cause of atmospheric electricity - the immersion of the cold upper strata of the atmosphere into the lower (warmer) layers causes mechanical friction of minuscular particles in the air against each other, that results in generation of atmospheric electricity. Two kinds of particles are required: those of water (vapors) capable of accumulation of the electricity, and other which are involved in production of the electricity via friction. The latter are organic compounds, which can not mix with water, "fatty substances…balls of flammable vapors… that appear in the air in a great variety from the body fumes of animals and humans", products of combustion, burning and rotting of all kinds of organics. Electrically charged droplets are assumed to be spread throughout the entire volume of the cloud. The transfer of charges from individual "fatty" particles to droplets of water in the clouds via countless collisions leads to formation in atmosphere, in clouds of strong electric fields, which are the cause of the appearance of lightning.

Today's explanation of the atmospheric electricity is much more complex – see e.g. [23] – but it involves many features of the Lomonosov theory and the vertical air movements as the centerpiece. Water (micro) droplets and ice crystals of various sizes are known to be the most important elements in the formation of the atmospheric electricity. Famous American atmospheric scientist Bernard Vonnegut commented in [24] "…It is worth recognizing earlier perceptions of convection in cumuli. Lomonosov (1753) was aware of updrafts and downdrafts and suggested that friction between them caused the electrification of clouds. Again, convection was proposed as the source of electrical energy, when Grenet (1947) in France published a novel theory of cumulus electrification in which a charge deposited on the upper surface of the cloud by electrical conduction was carried down to lower levels by upper-level downdrafts to accumulate and cause lightning." It is simply remarkable how close the illustrations of the evolution of the lightning clouds and cells depicted in Feynman's *Lectures on Physics* – see, e.g., figures on page 9-6 of volume II [25] – resemble Lomonosov's *Fig.2*. Involvement of organic or "fatty" particles was not confirmed in common lightning, but they been experimentally observed in a related phenomenon of ball-lightning [26]. On base of his original observations and measurements with Richmann's electrometer, Lomonosov concluded existence of electric fields in quiet atmosphere, i.e., not during a thunderstorm but in a clear, cloudless weather. Lomonosov was also the first to correctly proclaim the presence of the electricity-generating particles and processes all over the entire volume of a thundercloud, while until the end of the 19th century, it was commonly believed that the clouds are charged only over the surfaces.

Presentation and proof of the new concept of the atmospheric electricity take three quarters of the *Discourse*, the rest is dedicated to practical matters and expansion of the theory to other electrical phenomena. It was very appropriate for a public meeting to discuss countermeasures to mitigate the risks of lightning strikes. Notably, all of them were not Lomonosov's inventions, but instead were presented as kind of consensus among the experts. He listed three of them: hiding in underground facilities, especially those which have water above them (the method which is nor supported by any theory but supported by experience in Freiberg mines and in Japan, and sort of consistent with the notions of the water being an effective acceptor of electricity), the lightning rods with sharp edges and shaking the air. The latter two are presented without full certainty – "…could seemingly be successfully used". Such an attitude toward the lightning rods probably indicates that Franklin ideas were not yet fully accepted in Russia and in the European scientific circles actively communicating with the St.Petersburg Academy. The shaking of ether by church bell ringing and cannon firing was ideologically consistent with Lomonosov's views that "electric power" and lightning are due to vibrations of ether (illustrated in Lomonosov's *Figs. 12, 13* and *14*). At the same time it is remarkable that though his joint studies with Richman in April 1753 did not significant effect of firing dozens of firework canons during the Imperial Court celebration, the method was still considered as generally acceptable. Again, that might reflect nothing but common understanding of the times. For expample, eminent Dutch physicist, inventor of the first electric capacitor ("Leyden jar") Pieter van Musschenbroek (1692-1761) in the article on electricity for the famous French *Encyclopedie* (under "*Tonnerre*" [27]) stated that "…thunder can be disrupted and diverted by the sound of several bells or the firing of a cannon; in this way a great agitation is excited in the air, which disperses the parts of the lightning; but it is necessary to be careful not to ring when the cloud is precisely above the head, to avoid direct thunderbolt from the cloud splitting overhead." Musschenbroek defended such "traditional ways" of preventing lightning as quite effective in several other publications [28].

Lomonosov was not the only one in mid-18 century who stated the electrical nature Northern lights (aurora borealis). What was original in his approach is generation of the electricity from the movement of ascending and descending air currents in upper atmosphere in the polar regions – similar to what he proposed for the lightning – and the assertion that they ignite shining of the ether above the atmosphere – concluded out fundamental similarity of the auroras to a gas discharge in vacuum or thin gas. While the former is not true, the latter is correct. Observing aurora radiance in St.Petersburg on October 16, 1753, Lomonosov, as described in *Explanations*, was able to measure its height with an remarkable accuracy for those years and found the upper edge of the lights reaching about 420 versts, or 450 kilometers. That is compared to modern values of typical lower boundary of auroras at 95-100 km, and the upper edge between 400km to 600 km, as a rule, but sometimes up to 1000-1100 km.

Finally, Lomonosov briefly discusses the nature of comet tails, expresses doubts in the Newton's hypothesis [29] and proposes his own, naturally explaining the tails by electricity generation in the comet atmosphere down- and up-drafts at the borders of the shade areas (as in the theory of lightning) and by radiance of the electricity-induced vibrations of ether far beyond the atmosphere (i.e., like in his theory of Northern lights. None of these effects are involved in modern explanations of the phenomena of comet tails.

There are no documented evidences of the audience reaction to Lomonosov's presentation made on November 25, 1753, but present day reader of the *Discourse* should definitely be impressed by Lomonosov's wit, depth of his rational thinking and the breadth of his knowledge – he covered subjects and cited evidences from various epochs and from a wide range of sources, brought up interesting classification of various electrical phenomena in atmosphere, touched such diverse subjects as electro-mechanical responses of the mimosa plant [30], St.Elmo's lights and typhoons, he paid sincere tribute to his late colleague Georg Richmann and praised the support of sciences exhibited by Peter the Great and his daughter, governing Empress Elizaveta Petrovna.

In 1963, by B.Vonnegut's request, the American Meteorological Society commissioned David Krauss to translate *Discourse on Atmospheric Phenomena Originating from Electrical Force* from Russian to English. The draft manuscript with numerous hand-written corrections was never published and has been made available to author from the university archives at SUNY Albany, NY. Despite a number of mis-readings and mis-interpetations - very much excusable because of quite heavy kind of the 18[th] century Russian language of Lomonosov's *Discourse* - that draft has been widely used a reference for this translation.

I would like to thank Scott Vonnegut and Julia Whitehead of the Kurt Vonnegut Memorial Library (Indianapolis, IN), and Sally Marsh (formerly of the University of Albany) for their invaluable assistance with finding the 1963 English translation of the *Discourse on Atmospheric Phenomena Originating from Electrical Force by Mikhail Lomonosov* and Robert Crease of SUNY, my long-term collaborator and co-author of several scholar papers on Lomonosov, for help in understanding the history of the 18[th] century electricity research.